# Correlated Electron Materials and Field Effect Transistors for Logic: A Review

You Zhou and Shriram Ramanathan

*School of Engineering and Applied Sciences, Harvard University, Cambridge, Massachusetts 02138, USA*

**Abstract**

Correlated electron systems are among the centerpieces of modern condensed matter sciences, where many interesting physical phenomena, such as metal-insulator transition and high-$T_c$ superconductivity appear. Recent efforts have been focused on electrostatic doping of such materials to probe the underlying physics without introducing disorder as well as to build field-effect transistors that may complement conventional semiconductor metal-oxide-semiconductor field effect transistor (MOSFET) technology. This review focuses on metal-insulator transition mechanisms in correlated electron materials and three-terminal field effect devices utilizing such correlated oxides as the channel layer. We first describe how electron-disorder interaction, electron-phonon interaction and/or electron correlation in solids could modify the electronic properties of materials and lead to metal-insulator transitions. Then we analyze experimental efforts toward utilizing these transitions in field effect transistors and their underlying principles. It is pointed out that correlated electron systems show promise among these various materials displaying phase transitions for logic technologies. Furthermore, novel phenomena emerging from electronic correlation could enable new functionalities in field effect devices. We then briefly review unconventional electrostatic gating techniques, such as ionic liquid gating and ferroelectric gating, which enables ultra large carrier accumulation density in the correlated materials which could in turn lead to phase transitions. The review concludes with a brief discussion on the prospects and suggestions for future research directions in correlated oxide electronics for information processing.

## Contents

## I. Introduction

Field effect transistors are the building blocks of nanoelectronic devices and continuous innovations in this field have fueled the growth of semiconductor and related electronics industry for several decades. A key aspect is the geometric scaling of the channel length and associated critical dimensions which allows performance improvement at the circuit level while keeping power dissipation manageable. However, as the size of a single FET shrinks to the nanometer dimensions, it has become more and more difficult to turn OFF the transistor, or in other words, to reliably control the channel region conductance by the gate. At the very least, this could lead to additional power consumption due to undesirable leakage. This seemingly fundamental problem motivates the search for alternate approaches to computation ranging from materials replacements to novel architectures for logic operations.[1, 2]

Several efforts are on-going to critically evaluate prospects of alternate approaches for realizing the next generation switching devices for information processing. One approach is to continue shrinking the channel dimensions by developing field-effect transistors operating under different physical principles. The difficulty of controlling the OFF state current in short-channel conventional FET lies in the fact that the carriers in the channel are thermally activated and carrier density is controlled by the channel surface potential. Therefore, one could look for either a different conduction mechanism in the channel or a different coupling mechanism between the gate and channel to achieve better performance. For example, instead of using a gate oxide to fine-tune the channel conductance, using a ferroelectric material to change the channel conductance in two discrete levels could potentially consume less energy and increase the sharpness of the transconductance curve as suggested from modeling.[3] In a different type of device, tunnel field-effect transistor, the channel carriers are quantum-mechanically injected by band-to-band tunneling instead of by thermal injection.[4] Ultra-low power dissipation and operation voltage have been demonstrated in this design.[5] Another strategy is to introduce new functionality in FET devices. Traditionally FET devices are used as Boolean logic switches. One could think of implementing innovative logic operation and constructing new computational architecture based on devices with unusual behaviors. Neural computation that mimics the operation of biological systems could drastically increase parallel processing beyond Boolean computation.[6] There are also proposals of the usage of quantum computation that is superior to conventional computing algorithms for certain classes of problems.[7] In some of these ideas, spin instead of charge is used as the computational vector thanks to its long decoherence time and thus quantum information could be retained.[8-10]

Utilizing phase transitions in correlated oxides is another approach to explore switching action under gate bias in three-terminal geometry. Among materials showing metal-insulator transitions, there are a series of compounds in which their electronic phase diagram is coupled with the carrier densities, often referred to as Mott insulators. A certain amount of net carrier concentration induced by the gate could possibly lead to a sharp increase in its free carrier density thereby modulating the material resistance more dramatically compared with conventional FET. The subthreshold swing could potentially be smaller because of the enhanced conductance modulation. What is more, the phase transition could also give rise to new functionality that is not expected from conventional FETs. For example, carrier-mediated ferromagnetic or antiferromagnetic phases might be interesting for magnetic logic.

Figure 1(a) summarizes recent efforts to build three terminal field effect devices using different materials and/or mechanisms.[11-34] The first two rows show field effect transistors based on the traditional semiconductors. The first three devices are conventional MOSFETs and have similar subthreshold swing with a minimum of ~60mV/decade at room temperature, whereas the other five devices could have smaller subthreshold swing and may overcome the scaling limit of conventional MOSFETs. The last four rows show FETs based on 1-D structures, 2-D structures, organic materials and oxide materials, respectively. These devices are expected to have similar subthreshold behavior with conventional MOSFETs, except those based on correlated materials (Organic Mott FET, $SrTiO_3/LaAlO_3$ FET, $SrTiO_3$ and ionic liquid Mott FET in Figure 1(a)). Figure 1(b) shows the room temperature insulating phase resistivity and lattice parameters $a$ of some representative materials used as transistor components. Large resistivity is desired to achieve low off state current in FET operation. For materials such as Si[35] and Ge[36], the resistivity is taken from the wafer resistivity of demonstrated FET devices and of course depends to a great extent on the doping concentration. The OFF state channel resistance is larger than the estimation from wafer resistivity because of the *n-p-n* or *p-n-p* junctions in MOSFETs. For novel materials (2D crystals, Mott insulators etc.), their bulk resistivity values are plotted because FETs based on these materials are often fabricated from a homogeneously doped sample and the bulk resistivity is thus directly related to the OFF state current. For doped materials such as Nb-doped $SrTiO_3$ and $YBa_2Cu_3O_{7-x}$, resistivity could vary as a function of dopant concentration. Undoped $SrTiO_3$ and $KTaO_3$ have $d^0$ electron configuration and are therefore termed as 'band insulators', but correlation effects become important when they are doped with electrons.[37,38] For this reason, both of them are plotted as correlated materials. We could see from Figure 1(b) that several of the insulating correlated materials are not 'good' insulators due to their small band gap and/or defects formed during growth. Innovative device design is therefore crucial to achieve high-performance FET devices with these materials that do not consume substantial power in the OFF state. Lattice parameter provides important information for thin film growth as well as the estimation of the channel material's compatibility on relevant substrates. For example, the metal-insulator transition magnitude is degraded in $VO_2$ when it is deposited on lattice-mismatched or non-oriented substrates. Similarly, in several rare earth nickelates, it is extremely challenging to obtain phase pure material at ambient pressure in the absence of epitaxial strain. The strain from the substrate could tune the channel material properties, which is relevant to Mott FETs typically operating near the boundary of electronic phase diagrams. For many low symmetry materials such as $VO_2$ monoclinic phase, lattice constant $a$ may not be sufficient to estimate the film quality on a certain substrate. We summarized the lattice parameters and space group of these materials in Table I. Some of these correlated materials including rare earth nickelates, $La_2CuO_4$ and $La_{2-x}Sr_xCuO_4$ (LSCO) have distorted perovskite crystal structures and the corresponding thin films are often deposited epitaxial on cubic perovskite substrates. In the context of thin film applications, the pseudocubic lattice parameters are more frequently used to discuss film properties and included in the Table I(d). The materials that use pseudocubic lattice constant in Figure 1(b) and (c) are denoted by open symbols. Figure 1(c) shows the band gap of these materials. Comparing with Figure 1(b), it could be found that the insulating state resistance roughly increases with band gap. Exceptions could arise from the dopant concentration as well as the difference in carrier mobility in these materials. Note that many of the correlated materials such as some high $T_c$ superconducting cuprates are poor metals at room temperature and not shown in Figure 1(c). Also the band structures of some other correlated materials such as $SmNiO_3$ and $NdNiO_3$ are not yet well known or simply estimated.[39,40] Table I(d) summarizes the lattice parameters and space group of these materials. Some of the doped materials (for example, YBCO) may have various lattice parameters and space groups depending on the exact oxygen stoichiometry and doping concentration and only representative values are presented in the table.

In this paper, we review experimental progress towards building field-effect transistors (three-terminal devices) utilizing correlated electron metal-insulator transitions and also discuss the scientific underpinnings governing the switching mechanism. In section I, we give an introduction on different types of metal-insulator transitions with emphasis on the carrier transport properties. In section II, we review the operation mechanism of conventional FETs and define parameters that are essential for comparison to proposed ideas. In section III, we survey the demonstrated field-effect experimental work with each of these types of correlated materials. By looking into field-effect experiments in diverse correlated insulators, we emphasize that Mott transition FET is a particularly promising device. Section IV reviews technological progress towards building Mott FET with novel structures, such as ionic liquid and ferroelectric gating. Section V concludes with perspectives of comparison between conventional and alternate proposed technologies as well as future research needed to realize functional devices for information processing.

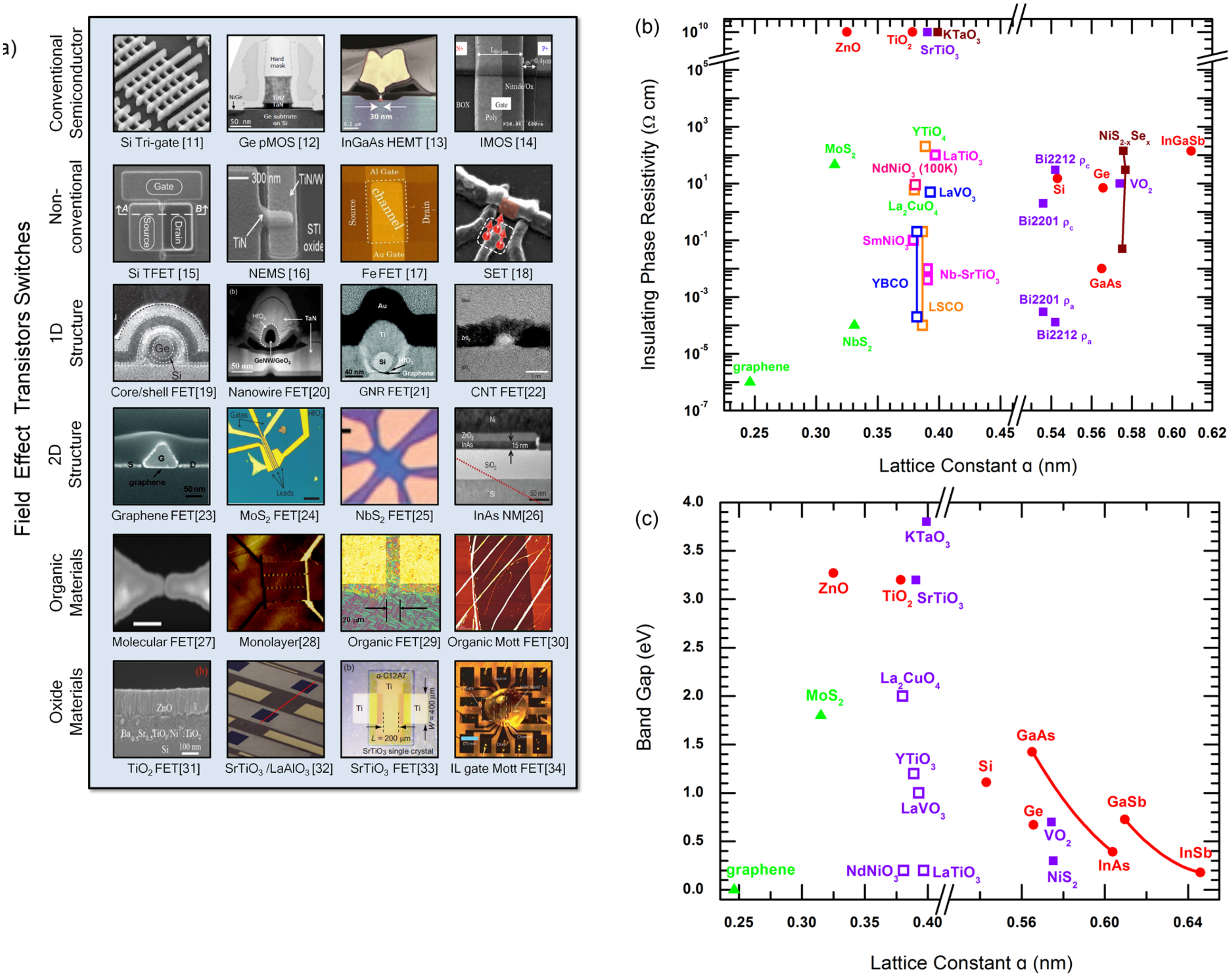

Figure 1. (a) Current efforts in building three-terminal field-effect devices using different materials and transition mechanisms. 1st Row (conventional semiconductors, from left to right): 22nm Tri-gate Si MOSFET[11], 65nm Ge pMOSFET[12], 30 nm $In_{0.7}Ga_{0.3}As$ Inverted-Type HEMTs[13] and impact ionization MOSFET (IMOS)[14]. 2nd Row (other operation mechanisms): SOI Si tunnel FET (TFET)[15], nano-electromechanical switch (NEMS)[16], ferroelectric FET (FeFET)[17] and single electron transistors (SET)[18]. 3rd Row (1D structure): Si/Ge Coreshell FET[19], Ge Nanowire FET[20], graphene nanoribbon (GNR) FET[21] and carbon nanotube (CNT) FET[22]. 4th Row (2D structure): graphene FET[23], $MoS_2$ FET [24], $NbS_2$ FET[25] and InAs nanomembrane FET[26]. 5th Row (organic materials): Molecular FET[27], organic monolayer FET[28], organic FET[29] and organic Mott FET[30]. 6th Row (oxide materials): $TiO_2$ FET[31], $SrTiO_3/LaAlO_3$ FET[32], $SrTiO_3$ FET[33] and ionic liquid (IL) gated Mott FET [34]. FET devices utilizing correlated electron systems include organic Mott FET, $SrTiO_3/LaAlO_3$ FET, $SrTiO_3$ FET and ionic liquid (IL) gated Mott FET and will be the focus of this article. Transistors based on operating mechanism different than conventional MOSFET could have different subthreshold behaviors. (b) Resistivity and lattice constant $a$ of various materials, including conventional semiconductors (circles), 2D crystals (triangles) and correlated materials (squares), which have been explored for FET applications (open symbols represent materials whose pseudocubic lattice constant is often used). Some of the resistivity values such as for Si[35], Ge[36] and GaAs[193] are extrapolated from the resistivity of wafer used in demonstrated FETs and of course depend on the doping concentration. The OFF state resistance of the channel could be larger than the estimation from wafer resistivity because of the *n-p-n* junction. Resistivity values of other materials are taken from values reported for bulk crystal or thin films.[194-201] For cuprates, the resistivity of single crystal is highly anisotropic: the resistivity in the $CuO_2$ plane ($\varrho_a$) is ~$10^{-3}$-$10^{-4}$ X the out-of-plane resistivity ($\varrho_c$) and also varies with doping concentration.[202-204] The lattice parameters are important to evaluate the material growth on certain substrates. ($TiO_2$ data for anatase phase) In addition, in some of the materials, for example, $VO_2$, the phase transition from insulator to metal is accompanied by a change in the lattice constant and point group symmetry. (c) Band gap of conventional semiconductors (circles), 2D crystals (triangles) and correlated materials[205-208] (squares) versus lattice constant $a$ (open symbols: pseudocubic lattice constant). The insulating state resistivity in (b) roughly increases with increasing band gap. Some of the cuprates in (b) are metallic[202] and not shown in (c). For correlated materials including $NdNiO_3$ and $SmNiO_3$, the knowledge of band structure is not well-established[40]. The band gap of III-V compound ternary alloys such as GaInAs and GaInSb are calculated using equation from [209]. Also note that the band gaps of 2D monolayers are typically different from their bulk values.[210] (Device images in panel a adapted with permission from references 11-34, respectively.)

**Table I(d)** Lattice parameters and space group of materials plotted in Figure 1(b)

| Material | Lattice constant a (nm) | Lattice constant b (nm) | Lattice constant c (nm) | Pseudo-cubic lattice constant a (nm) | Space group |
|---|---|---|---|---|---|
| Si | 0.5431 | 0.5431 | 0.5431 | | $F d \bar{3} m$ |
| Ge | 0.5657 | 0.5657 | 0.5657 | | $F d \bar{3} m$ |
| GaAs | 0.5654 | 0.5654 | 0.5654 | | $F \bar{4} 3 m$ |
| InAs | 0.6036 | 0.6036 | 0.6036 | | $F \bar{4} 3 m$ |
| InSb | 0.6461 | 0.6461 | 0.6461 | | $F \bar{4} 3 m$ |
| GaSb | 0.6118 | 0.6118 | 0.6118 | | $F \bar{4} 3 m$ |
| $SrTiO_3$ | 0.3905 | 0.3905 | 0.3905 | | $P m \bar{3} m$ |
| $KTaO_3$ | 0.3983 | 0.3983 | 0.3983 | | $P m \bar{3} m$ |
| $LaTiO_3$ | 0.5633 | 0.5617 | 0.7915 | 0.3968 | P n m a |
| $YTiO_3$ | 0.5316 | 0.5679 | 0.7611 | 0.3890 | P n m a |
| $LaVO_3$ | 0.5558 | 0.553 | 0.7849 | 0.3928 | P n m a |
| $SmNiO_3$ | 0.5328 | 0.5437 | 0.7567 | 0.3795 | P n m a |
| $NdNiO_3$ | 0.5382 | 0.5386 | 0.7360 | 0.3810 | P n m a |
| $La_2CuO_4$ | 0.5356 | 0.5405 | 1.3143 | 0.3805 | C m c e |
| $La_{2-x}Sr_xCuO_4$ | 0.3865 | 0.3865 | 1.9887 | 0.3865 | I 4/m m m |
| YBCO | 0.386 | 0.388 | 1.168 | 0.386 | P m m m |
| Bi2212 | 0.5406 | 0.5406 | 3.0715 | | A m a a |
| Bi2201 | 0.5469 | 0.5483 | 0.5426 | | F m m m |
| $VO_2$ (M1) | 0.5743 | 0.4517 | 0.5375 | | $P2_1/c$ |
| ZnO (Wurtzite) | 0.3249 | 0.3249 | 0.5205 | | $P\ 6_3\ m\ c$ |
| $TiO_2$ (Anatase) | 0.37845 | 0.37845 | 0.95143 | | $I\ 4_1/a\ m\ d$ |
| $NiS_2$ | 0.5676 | 0.5676 | 0.5676 | | $P a \bar{3}$ |
| NiSe | 0.3661 | 0.3661 | 0.5356 | | $P\ 6_3/m\ m\ c$ |
| $MoS_2$ | 0.316 | 0.316 | | | $P\ 6_3/m\ m\ c$ |
| $NbS_2$ | 0.331 | 0.331 | | | $P\ 6_3/m\ m\ c$ |
| graphene | 0.2461 | 0.2461 | | | P 6/m m m |

In literature, the notation 'P b n m' is often used instead of 'P n m a' (standard).

## II. Types Of Metal-Insulator Transitions

The simple single electron band picture describes the electronic structure of a solid by considering electrons moving independently in an effective periodic potential imposed by ions and other electrons. The periodic potential makes electron dispersion curves discontinues at Brillouin zone boundaries and forms energy bands and gaps in the energy space. Metals contain a partially filled band while insulators' electron bands are fully occupied at zero temperature. Electrons move freely in the periodic potential and are scattered by impurities in the lattice. This picture works exceptionally well in many materials, but it neglects disorder-electron interaction, phonon-electron interaction and electron-electron interaction or treats them as perturbations. Consequently, in some scenarios where these interactions become strong, it is expected that the above approximation would fail to provide a full account of corresponding physical phenomena.

Experimentally, the deficiency of simple band picture is especially pronounced when studying a series of materials that show a transition from insulating to metallic state under certain conditions. In these materials, the aforementioned interactions largely overlooked in single-electron band picture become important and we can in fact categorize the metal-insulator transitions based on the dominant interaction that drives the material into the insulating phase: disorder-electron interaction, lattice-electron interaction and electron-electron interaction could induce Anderson localization, Peierls transition and/or Mott transition, respectively.

### A. Anderson localization

Lattice disorder such as impurities and vacancies could scatter conducting electrons and decrease their mean free path and thereby material conductivity. Anderson[41] considered the limit of strong disorder at low temperatures without thermal excitations and found that instead of a slow decrease, electron diffusion completely stops beyond some critical disorder density. As a result, the conductivity would vanish to zero rather than go to a minimum conductivity. We can examine the effects of disorder based on the tight binding model. Consider each lattice site as a potential well that confines electrons at low temperatures. When no disorder is present, each site has the same set of eigen wavefunction and eigen energy and electrons can tunnel through the barriers with their energy levels aligned. Now, if strong disorder is introduced into the lattice by letting the well depth to fluctuate randomly, the eigen energy of each isolated site will become different. From first order perturbation theory, the portion of eigen wavefunction mixing is roughly inversely proportional to the difference between their eigen energies. Thus, in the limit of strong disorder, the eigen wavefunction will be dominated by a single site wavefunction instead of being a uniform mixture of wavefunctions on all lattice sites. In other words, electron eigen wavefunctions become localized wave packets and the conduction of the material would vanish at low temperature as shown in Figure 2(a) and (b).[42] It can be seen that the origin of Anderson localization is due to wave-like properties of electrons. The interference between different electron multiple scattering paths makes the probability of backscattering higher than expected from classical mechanics.[43] The enhancement in probability of backscattering is also known as weak localization.

For quantum states of a given energy in an disordered system, they are either all localized or all delocalized.[44] Consequently, there may exist a transition between the localized and metallic phases. Mott[45,46] developed the above idea into the notion of mobility edge, which is an energy level that separates extended and localized states as shown in Figure 2(b). Electrons with energy above the mobility edge have extended wavefunction and conduct current, whereas electrons near or below the mobility edge are localized[42]. A metal-insulator transition could be induced when there is a carrier density change in the extended states above mobility edge. Theoretical studies show that in 1D and 2D systems, any amount of disorder could lead to electron localization, therefore a real Anderson transition is not achievable. But in 3D systems, the electrons could be in either extended or localized states and weak localization could happen before strong localization occurs[43].

One thing to notice is that the localized states may still appear to be metallic, if the system size is smaller than the size of localized wave packet. Therefore it could be expected that the conductance of an Anderson insulator $g$ will change with varying the sample size $L$ while keeping the disorder density unaltered. Abrahams et. al.[47] studied how the conductance $g$ scales with system size $L$ and developed a scaling theory of localization as shown in Figure 2(c). It defines a scaling function $\beta(g) = \frac{d\ln(g)}{d\ln(L)}$ and predicts a size-dependent diffusion/conduction for electrons. For example, $\beta(g)$ is always negative in 1D and 2D disordered systems, so they may conduct current but conduction decreases with increasing system size and will eventually vanish in large systems. For 3D systems, the scaling theory predicts a critical point, at which point $\beta(g)$ changes sign and this is related to the mobility edge.

Experimentally, electron localization has been studied in various highly doped semiconductors, including Sb:Ge[48], Si:P[49,50], and Si:B[51]. But localization due to disorder is non-trivial to confirm because the Coulomb interaction between electrons also affects conduction behavior and complicates the data interpretation. The idea of Anderson localization, however, has been tested to be valid in many other systems such as light,[52] sound waves[53] and matter waves in Bose-Einstein condensate.[54]

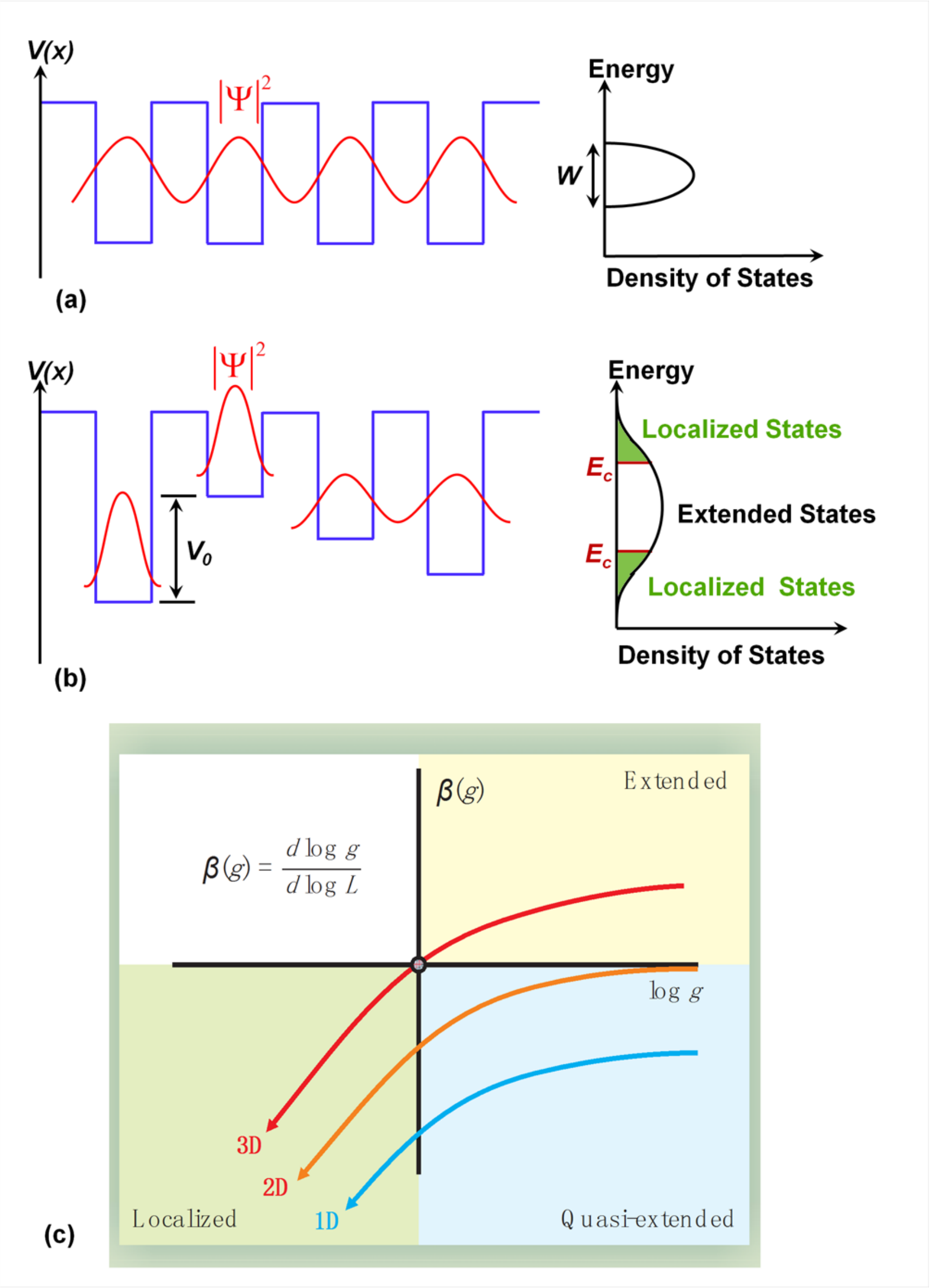

Figure 2. The lattice potential and density of states of (a) a perfect crystal and (b) a disordered crystal. For electronic states with certain energy in a disordered system, they should be either all extended or all localized. Thus there would be a 'mobility edge' $E_C$ that separates the localized state from extended state in the density of states diagram. A phase transition could happen if Fermi energy moves above or below the mobility edge. (c) The conductance of an Anderson insulator $g$ depends on the size of the sample. In the scaling theory, $\beta(g) = \dfrac{d\ln(g)}{d\ln(L)}$ describes how $g$ changes in accordance with the size $L$. For localized states, the conductance decreases with sample size and $\beta(g)$ is negative $L$, while the conductance increases with $L$ for extended states. The value of $\beta(g)$ depends on the dimension of the insulator and is always negative for 1D and 2D cases. In 3D systems, the critical point where $\beta(g)$ changes sign corresponds to the phase transition and mobility edge. (Panel a adapted with permission from ref [42]. panel c reprinted with permission from Lagendijk et al. [43] Copyright 2009, American Institute of Physics.)

## B. Peierls transition

In 1955, R. Peierls predicted that 1D metals could not exist at zero temperature.[55] Suppose a one-dimensional metal that is composed of equally spaced atoms with lattice constant *a* and the conduction band is partially filled as shown in Figure 3(a), where $k_F$ is the Fermi wave vector of the metal. If the position of atoms and equally the lattice potential are periodically modulated and the wavelength of such modulation is $\lambda_c = \pi/k_F$ (correspondingly wave vector Q = $2k_F$), the boundary of Brillouin zone will coincide with the Fermi surface and opens a band gap at the Fermi surface (Fermi surface nesting). The size of the crystal unit cell will become identical to the wavelength $\lambda_c$ and such periodic modulation of atom positions is also known as charge-density-wave (CDW).[56] The spatial periodicity $\lambda_c$ could be either commensurate or incommensurate with original lattice constant *a*. For 1D systems, the band openings lead to a reduction in the energy of electrons near the Fermi surface as shown in Figure 3(b), while the crystal elastic energy increases in order to deform the lattice. Peierls showed that the energy reduction from electrons is always larger than the gain in the elastic energy at T = 0K.[57] As a consequence 1D metals are susceptible to the formation of CDW and could not exist at low termperatures. At non-zero temperature, the energy reduction from band gap opening becomes smaller due to Fermi distribution and there would be a critical temperature where a Peierls insulator would transit into a metal. In three-dimensions, however, the Fermi surface is usually a sphere or part of a sphere and consequently it is almost impossible to have a periodic deformation that could lead to exact Fermi surface nesting. But in materials with highly anisotropic and quasi-one-dimensional band structures, such as $NbSe_3$[58], $K_{0.03}MoO_3$[59] and organic conductors TTF-TCNQ[60], the formation of CDW ground states and Peierls transition have been verified by temperature dependent transport phenomena as well as superlattice reflections. Because many of these Peierls insulators, such as $NbSe_3$, $NbSe_2$ and $NbS_2$ (transition metal dichalcogenides), are quasi-2D structures with strong in-plane bonding and relatively weak out-of-plane bonding, they could be readily cleaved into 2D monolayers, which is gaining increasing interests.[25,61]

The conduction mechanism in Peierls insulators is different from band insulators. Besides electron excitation across the band gap, the collective motion of CDWs will also contribute to the conduction[56]. Due to the presence of impurities in the crystal, CDWs will have preferred position relative to the crystal to minimize the energy and are hence pinned to the underlying lattice. However, under a threshold electric field $E_T$, the CDWs could become de-pinned and slide in the crystal. The sliding of CDWs does not conduct a current by itself since the motion of ions are periodic, but it will modulate the crystal potential, which could couple with the electrons and form a current in Peierls insulators. Upon reaching the electric field $E_T$, the current increases dramatically as shown in Figure 3(c)[56] and as large as ~9 orders of magnitude change in current has been reported.[56,62,63] Peierls insulators are also called charge-density-wave conductors based on the conduction mechanism. Another interesting phenomenon in Peierls insulators is that the CDW's response to a given pulse will depend on the history of operation.[64,65] This memory effect is caused by different metastable states of CDW (pinned on different impurity sites) in the crystal and may be of technological interest for 'memory resistors'. Detailed dynamics of CDWs and other experimental phenomena could be found in other review papers.[56, 66]

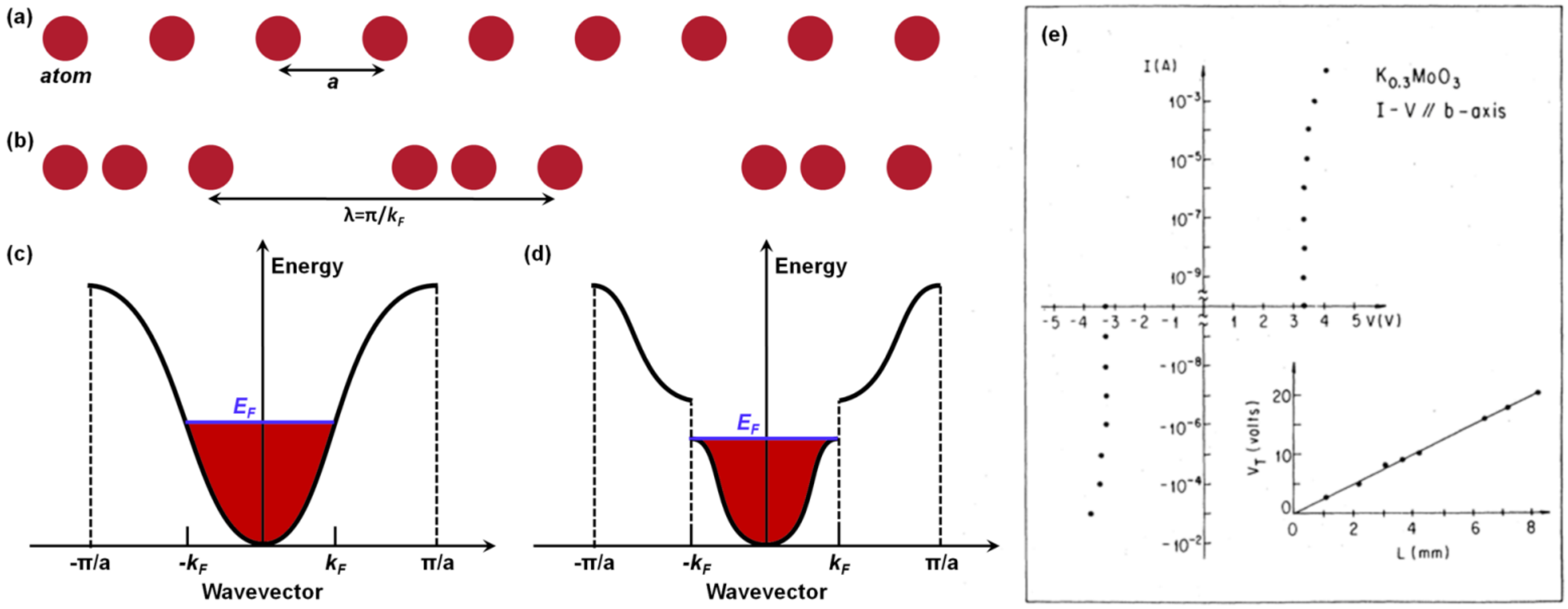

Figure 3. The atomic configuration of (a) a one-dimensional normal metal and (b) a Peierls insulator and their band structure (c) and (d). The 1D metal is susceptible to the formation of a periodic modulation of atom positions and thereby electron densities. The modulation opens up a band gap at the Fermi surface and makes the material an insulator. The insulator (b) is energetically favorable than undistorted metal (a) at low temperatures. At high temperatures, the undistorted metal becomes the stable phase and a insulator-to-metal transition is induced. (e) In Peierls insulators, the periodic modulation (charge-density-wave) could move under an external electric field and contribute to the sample conduction. The current-voltage characteristics of $K_{0.3}MoO_3$ indicate two conduction mechanisms at low temperatures. Below certain threshold voltage $V_T$, the conduction is dominated by the free carrier thermally excited across the Peierls band gap. Beyond $V_T$, a small change in external voltage could lead to magnitude change in the current because of the motion of charge-density-wave. (Panels a to d reprinted with permission from Thorne. [66] Copyright 1999, American Institute of Physics; panel e adapted with permission from Grunner. [56])

## C. Mott transition

Coulomb repulsion could localize electrons which happens in Mott insulators.[67] The Mott insulating states could be understood from a tight-binding lattice model with *N* primitive unit cell and one electron on each site. In the simple band theory, the band degeneracy would be *2N* by taking into account of spin degeneracy and the material is metallic with half-filled conduction band – however far apart the atoms are. This is clearly inconsistent with the fact that isolated atom arrays will be not conducting as firstly pointed out by Mott.[68] It is now clear that the simple band picture is not sufficient because it neglects the Coulomb interaction between electrons, the so-called 'electron correlation'. For example, in the above tight-binding model, conduction happens by electrons hopping through the lattice from one site to another. When the electron is to hop from its original site to a new site, which is already occupied by another electron, it will experience Coulomb repulsion from the electron in the new site. If the Coulomb repulsion energy *U* or blockade is much larger than electrons' kinetic energy, electrons would be bounded to their original sites instead of being itinerant in the lattice. This effectively splits the original single half-filling band into a full lower band (lower Hubbard band, LHB) formed from electrons occupying an empty site and an empty higher band (upper Hubbard band, UHB) from electrons occupying a site already with one electron as shown in Figure 4(a). The magnitude of band splitting is the Coulomb repulsion energy *U*. This is the well known Hubbard model[69] and insulators with strong electron correlations are often referred to as Mott insulators. Typical Mott insulators include many of those transition metal (*4d-*, *4f-*, *5f-*) oxides (TMOs). According to band theory, most of these TMOs would be metals with partially filled *d* or *f* bands. However, many are insulators due to electron correlation and the transition metal *d* or *f* band splits into lower and upper Hubbard band. The dominant conducting mechanism is due to *d* electrons hopping between different transition metal atoms and could be described as[70]

$$d^n + d^n \rightarrow d^{n+1} + d^{n-1} \quad (1)$$

in *3d-* TMOs, where $d^n$ denotes an transition metal atom with *n d* electrons. Thus the Coulomb repulsion energy is

$$U = E(d^{n+1}) + E(d^{n-1}) - 2E(d^n) \quad (2)$$

where $E(d^n)$ is the total atomic energy with electron configuration of $d^n$. The Hamiltonian of the Hubbard model could therefore be formulated as,

$$H = -t\sum_{<ij>}\left(c_{i\sigma}^{+}c_{j\sigma} + \text{H.c.}\right) + U\sum_{i} n_{i\uparrow}n_{i\downarrow} - \mu\sum_{i\sigma} n_{i\sigma} \quad (3)$$

where $c_{i\sigma}^{+}$ ($c_{i\sigma}$) is the creation (annihilation) operator of single electron on site *i* with spin polarization $\sigma$, $n_{i\sigma} \equiv c_{i\sigma}^{+}c_{i\sigma}$ is electron number operator for the corresponding state, *t* is electron kinetic energy, *U* is the intrasite Coulomb repulsion energy and $\mu$ is the chemical potential.

In the limit of large Coulomb repulsion *U*, Hubbard model gives an insulator. With a fixed number of total electron density, reducing Coulomb repulsion *U* to a certain critical value could merge two Hubbard bands and induce an insulator to metal transition. The density of states with varying relative magnitude of Coulomb repulsion to bandwidth *W* is shown in Figure 4(b). For example, in transition metal oxides, the Coulomb repulsion energy *U* is usually a few eV (~1 to 10 eV),[70] whereas the bandwidth of *d* band are often quite small (~1 to 2 eV) because of the tight binding nature of *d* orbitals.[71] Notice that for conventional semiconductors such as Si and Ge, the bandwidth is usually much larger ~10 - 20 eV[72] with similar on-site Coulomb repulsion *U* to TMOs[73], which gives a hint on why correlation effect is not as important in conventional semiconductors.

While the above band closing picture of insulator-to-metal transition could reflect many aspects of the insulating side of the MIT because it starts from the limit of large *U*, it does not provide a good description of the metallic state.[74] Using the same Hubbard Hamiltonian, Brinkman and Rice[75] began their treatment of metal-insulator transition from the metallic side using Gutzwiller's variational method.[76] The theory models the metallic phase as a strongly renormalized Fermi liquid with renormalized Fermi energy $\varepsilon_F{}^*$. The renormalized Fermi energy decreases as the Coulomb repulsion $U$ increases and finally vanishes at a critical interaction value $U_c$. The effective mass of quasiparticle diverges as $m^* \propto \left(1-U/U_c\right)^{-1}$, when the electron-electron correlation approaches $U_c$. The metal-insulator transition is driven by the disappearance and mass-divergence of quasi-particle in the Brinkman-Rice scenario. This is different from the metal-insulator transition by vanishing carrier density $n$ in the above band closing picture and could be important for the analysis of experimental data.

In real materials exhibiting Mott transition, the problem is further complicated by intersite Coulomb repulsion, orbital degree of freedom, exchange interaction and spatial inhomogeneity in the material. Theoretical treatments of Mott transition could be found in other references.[77] Although various Hamiltonians are used in the theoretical treatment, the picture of band splitting could be generally applied for the insulating phase.

In addition, for many of these compounds, the influence of oxygen orbitals is not negligible. Assuming there is no hybridization between transition metal *3d* band and oxygen *2p* band, one could categorize TMOs into two different types based on the relative level of O *2p* band:[78] Mott-Hubbard type, where O *2p* lies in between two Hubbard bands, and charge-transfer type, where O *2p* level is under the lower Hubbard band as drawn in Figure 4(c) and (d). In the real space picture, conducting electrons hop through transition metals in Mott-Hubbard insulator whereas hopping occurs between metal and oxygen atom in a charge-transfer insulator.

The Mott metal-insulator transitions discussed heretofore are all induced by tuning relative magnitude of the Coulomb repulsion $U$ to bandwidth $W$ at fixed band-filling (half-filling), which is called bandwidth controlled metal-insulator transition. One example of this is applying stress on a Mott insulator to change the atom spacing and consequently bandwidth to induce a metal-insulator transition without changing the carrier density.

The other way to induce a metal-insulator transition in a correlated electron system is to change the band-filling, or in other words, to dope the upper (lower) Hubbard band with holes (electrons). Figure 4(e) and (f) illustrates how carrier doping could induce metallic phase in a Mott insulator.[79] At exact half-filling, electrons cannot hop to another site due to Coulomb repulsion and all the sites are singly occupied. If holes are doped into the insulator, however, some atom sites become unoccupied and the nearby electrons could move freely onto this unoccupied site, because the total energy is the same before and after the hopping. The original insulator now becomes metallic. Similarly, doping electrons into the insulator also makes electrons to hop freely through those unoccupied atom sites. The above metal-insulator transition controlled by carrier density is called band-filling controlled metal-insulator transition.

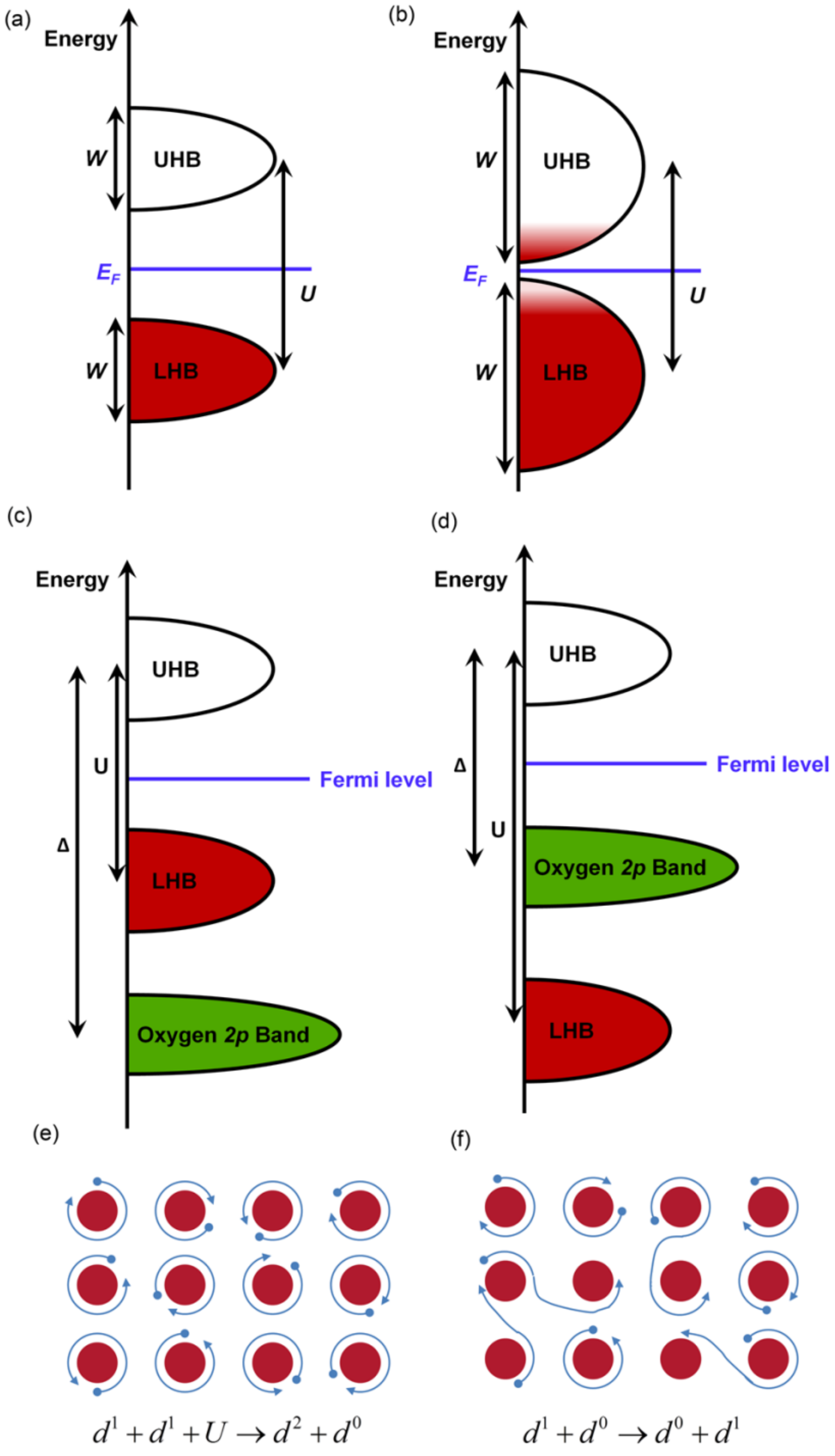

Figure 4. (a) Splitting of a normal band into upper and lower Hubbard band due to electron correlations. (b) Bandwidth controlled metal-insulator transition. Changing the bandwidth *W* could induce a metal-insulator transition. (c) and (d) Oxide Mott insulators could be categorized into two types based on the relative position of the oxygen band and Hubbard bands[78]: (c) Mott-Hubbard insulator where the oxygen p-band lies under the lower Hubbard band. (d) Charge-transfer insulator where oxygen p-band is in between the lower Hubbard band and upper Hubbard band. (e) and (f) Doping holes or electrons into a Mott insulator could lead to a phase transition from insulator to metal. (e) In an undoped Mott insulator (n=1), electron hopping leads to formation of a doubly occupied site and increase in total energy because of the Coulomb repulsion between two electrons. (f) In a Mott insulator with hole doping (n<1), electron hopping does not create doubly occupied sites and system total energy does not change. Electrons could move freely in the matrix and the material becomes metallic. (Panels c and d adapted with permission from Zaanen et al. [78]; panels e and f adapted with permission from Fujimori. [79])

One might expect that any finite carrier doping could make the Mott insulators conductive, because the doped electrons/holes can move freely and contribute to metallic conduction. What is often observed in experiments, however, is that a certain critical carrier density is requried before Mott insulators become metallic. Figure 5(a)-(c) shows the phase diagram of some typical Mott insulators.[77,80,81] Typically 10%-30% carrier doping per unit cell may induce a phase transition.[77] Mott addressed the problem of doping induced metal-insulator transition by considering the Coulomb potential screened by free electrons and proposed a criterion for the transition to occur:[82]

$$n_c^{1/3} a_H \sim 0.25 \qquad (4)$$

where $n_c$ is critical carrier density at the metal-insulator transition at T=0K, $a_H$ is the Bohr radius of electrons orbiting around the dopant center. Mott criterion seems to be an effective indicator of the critical condition for metal-insulator transition at finite temperatures.[83] Though originally developed for doped semiconductors, this 'simple' criterion turns out to be quite universal such that metal-ammonia, metal-noble gas systems and superconducting cuprates follow the relation as shown in Figure 5(d).[83-85] The validity of Mott criterion for such a variety of systems (especially for cuprates) is remarkable because of the simplicity of this original physical model.

Another interesting phenomenon as inferred from Figure 5(a) is that the hole doping and electron doping are asymmetric in Mott insulators. For example, in cuprate superconductors, the insulating state at half filling is more susceptible to hole doping than to electron doping. How the density of states $\rho(E)$ evolves with varying carrier doping is essential towards the understanding of band-filling controlled metal-insulator transition and would be further discussed in later section.

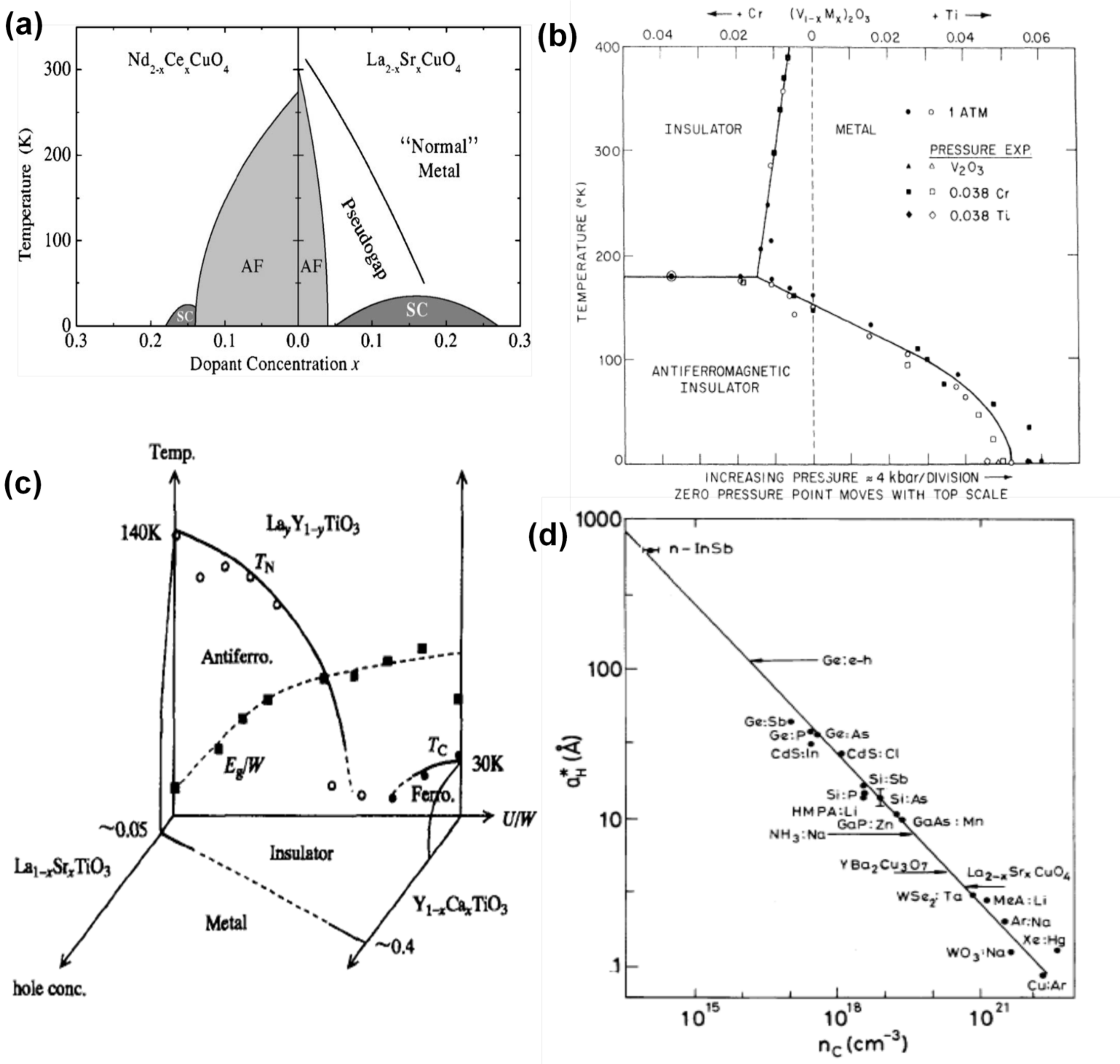

Figure 5. (a) Phase diagram of a cuprate superconductor[80]. A transition from antiferromagnetic insulator to metal could be induced near room temperature by carrier doping. The phase diagram is not asymmetric between electron and hole doping. (b) Phase diagram of a prototype Mott insulator, $V_2O_3$ as a function of both pressure and alloying[81]. The bottom x axis is pressure whereas top x axis shows the doping concentration. (c) Phase diagram of $R_{1-x}A_xTiO_3$ (R= La, Y. A = Sr, Ca).[77] Doping with same valence state atoms (La or Y) controls the band width/Coulomb repulsion ratio, while doping with different valence state substitutions alters the band filling. (d) The relation between metal-insulator transition critical density and dopant Bohr radius in various materials.[83] (Panel a adapted with permission from Damascelli et al.[80]; Panel b adapted with permission from McWhan et al.[81]; Panel c adapted with permission from Imada et al.[77]; Panel d adapted with permission from Edwards et al.[83])

## II. Conventional MOSFETs

### A. Operation mechanism

MOSFETs are primarily used as logic switches albeit they can also be employed in a myriad of other applications, which particularly requires low-voltage, low-power and high-speed operation. In this section, we briefly review the challenges facing Si-based MOSFET technology and introduce some terminology that could be helpful for understanding field-effect phenomena in other materials.

Figure 6(a) show the structure of classical n-channel MOSFET, or NMOS. In this device, two n-type doped regions (source and drain) are separated by a p-type semiconducting region (channel). A thin layer of insulating material (gate oxide) such as silicon dioxide or hafnia covers the channel area and a metallic gate electrode is deposited on top of the gate oxide. The source and substrate are typically grounded while varying gate voltage $V_G$ and drain to source voltage $V_{DS}$ in the MOSFET operation. For an ideal MOSFET, when no gate bias is applied on the gate electrode, the drain-channel-source acts as two back-to-back p-n junctions and there is almost no current flow across drain and source, which is the "OFF" state. At a sufficiently large positive gate bias, electrons are attracted to the gate oxide/channel interface and the dominating carrier type is inverted to electrons within the channel, which forms an n-type thin conducting channel connecting the source and drain. A current ($I_D$) could hence flow from drain through the source and the MOSFET is turned "ON" by the gate bias. The gate voltage could therefore modulate the channel conductance through field-effect. The voltage at which point inversion layer begins to form is referred to threshold voltage $V_T$. $I_D$ increases exponentially with $V_G$ when $V_G$ is approaching threshold voltage and this is called 'subthreshold region'. Above the threshold voltage, the drain-source current increases linearly with increasing $V_G$ at a fixed small drain bias. For small drain voltage, the drain source current is proportional to $V_{DS}$ and it is called linear region. Larger drain voltage could lead to pinch off and drain-source current saturation. Note that it is possible to replace the differently doped areas by a homogeneous doped Si and modulate the conductance by gate voltage, but it is difficult to achieve high ON/OFF ratio in such conventional MOSFET structure because Si is not a 'good' insulator and the gate can only control the conductance of a thin layer.

### B. Device parameters and scaling limit of MOSFET

The gate voltage only penetrates upto a thin layer and modulates the carrier density within this length (Debye screening length $L_D$) because the external field is screened by free carriers. For non-degenerate statistics, $L_D$ is given by

$$L_D = \left( \frac{\varepsilon_r \varepsilon_0 k_B T}{e^2 (p+n)} \right)^{\frac{1}{2}} \tag{5}$$

where $\varepsilon_r$ is the dielectric constant of the channel material, $\varepsilon_0$ is the vacuum permittivity, $k_B$ is the Boltzmann constant, $T$ is temperature, $n$ and $p$ is electron and hole density, respectively. Because $V_G$ will drop on both gate oxide and the screening layer in the semiconductor, the differential capacitance of the MOS system could be treated as gate oxide capacitance $C_{ox}$ in series with surface charge capacitance $C_{Si}$. The potential drop will be divided between gate oxide and channel in proportion to the inverse of their capacitance.

For an ideal logic device, the $I_D$ versus $V_G$ curve should be as sharp as possible, i.e., the drain-source current should be zero below threshold voltage (subthreshold region) and become a finite value above it. However, in MOSFET the $I_D$ is non-zero but increases exponentially with $V_G$ below $V_T$ as shown in Figure 6(b). The term to measure the steepness of the transition from OFF to ON is the subthreshold swing, *S*. Subthreshold swing indicates how much gate voltage needs to be increased in the subthreshold region to enhance the drain source current by tenfold. For conventional MOSFET,

$$S \equiv \frac{dV_G}{d(\log_{10} I_D)} = \frac{dV_G}{d\Psi_s}\frac{d\Psi_s}{d(\log_{10} I_D)} \cong \left(1+\frac{C_d}{C_{ox}}\right)\frac{k_B T}{q}\ln 10 \qquad (6)$$

where $\Psi_s$ is the surface potential, $C_d$ is the capacitance of depletion layer ($C_{Si}$) in the subthreshold region and $q$ is the electron charge. As previously seen, $V_G$ will be divided between the gate oxide and channel and $1+\frac{C_d}{C_{ox}}$ (referred to as the body factor *m*) gives how efficient semiconductor surface potential is coupled to the gate voltage. The body factor *m* is typically 1.2 to 1.5 for conventional MOSFET.[86] The value of $\frac{d\Psi_s}{d(\log_{10} I_D)}$ depends on the operation mechanism of the FET and for a conventional MOSFET, it is given by $\frac{k_B T}{q}\ln 10$, which is ~60 mV at 300K. This is because the inversion charge density is roughly proportional to $e^{q\Psi_s/k_B T}$ in the subthreshold region. In general, no matter what channel material (Si, Ge, III-V and so on), gate oxide design (high-ϰ dielectrics, multiple gates and other) are being utilized in MOSFET, as long as the same physical principle is used, the subthreshold swing has the minimum value of ~60 mV decade$^{-1}$ at room temperature. This provides crucial insight into a fundamental limit of MOSFET scalability as will be seen later.

One way to increase FET device switching speed is to enhance the carrier mobility in the channel. Thus the carrier mobility is an important parameter for FET operation. In many cases, it is more convenient to define, the field effect mobility, often measured above $V_T$ in linear region, as

$$\mu_{FE} = \frac{\partial I_D}{\partial V_G}\frac{L}{C_{ox} W V_{DS}} \qquad (7)$$

where *L* is the channel length and *W* is the channel width. The field-effect mobility is not the same but has similar values to the carrier mobility.

Another path towards faster operation of FETs is to shorten the channel length. The shrinkage of a single MOSFET's dimension (a feature is the channel length *L*) has fueled the development of VLSI, not only enhancing transistor density and speed, but also achieving lower power dissipation and lower cost per transistor. However, as previously seen, subthreshold swing is non-scaling and thus a minimum gate voltage swing is required to sustain a desired ON/OFF ratio. The OFF current will increase significantly with decreasing $V_T$, which sets a lower limit on the threshold voltage $V_T$ as well as power-supply voltage $V_{DD}$. Constant-voltage scaling is used to improve the transistor density on chip, but this will also lead to increase in OFF current even if the threshold voltage is unchanged due to short channel effect such as drain-induced-barrier-lowering (DIBL), which causes $V_T$ to decrease with increasing $V_{DD}$.[87] Although some technologies like silicon on oxide (SOI) and multigate structure may provide solutions for the next decade[88], the energy dissipation in these devices could be enormous. Therefore logic devices operated with novel physical principles might be an alternative toward more compact, faster or more energy efficient computing. In the following, we examine some examples that attempt to utilize the metal-insulator transitions in a FET design and speculate on their possible technological applications.

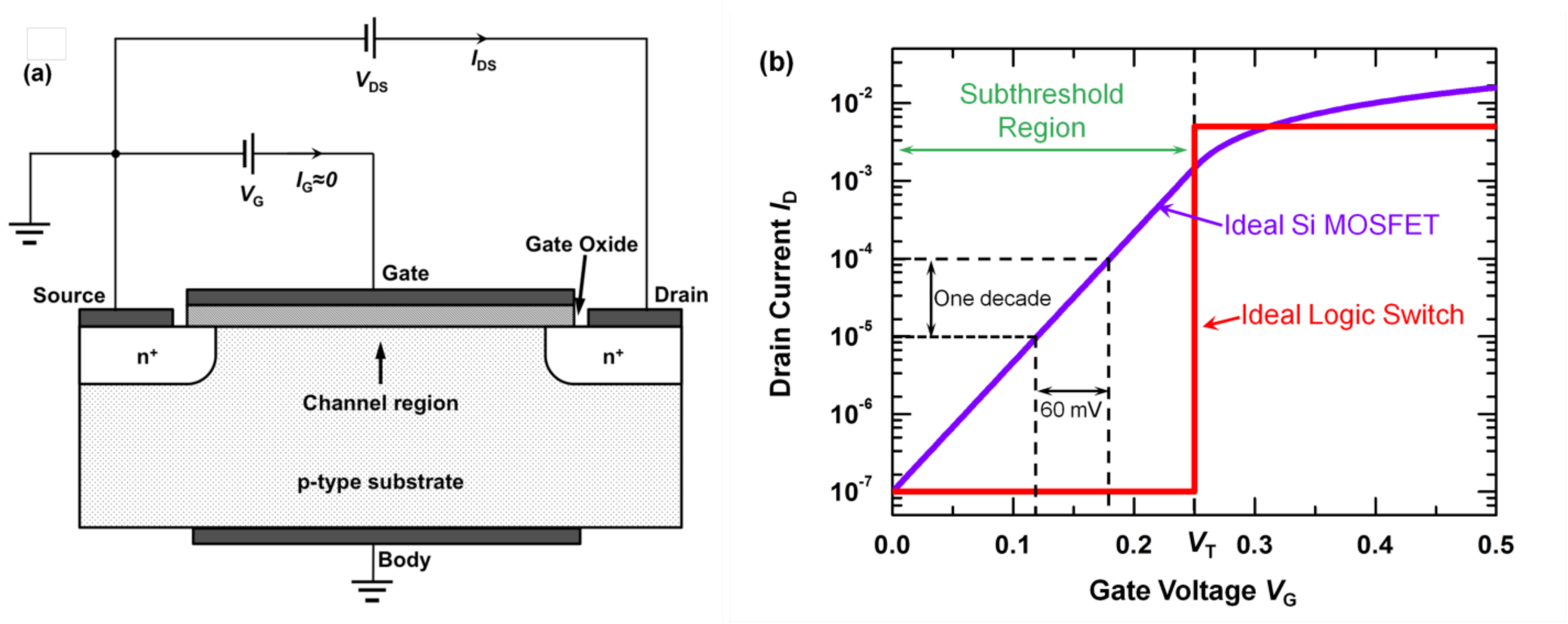

Figure 6. (a) Schematic structure of a conventional metal–oxide–semiconductor field-effect transistor (MOSFET). Two n-type regions (channel and source) are separated by a p-type channel. A gate insulator and metal electrode is deposited on the channel. When a positive bias is applied to the gate, the channel becomes n-type and the channel conductance increases. (b) The subthreshold behavior of an ideal logic device and ideal conventional FET device. The subthreshold swing for conventional FET is ~60 mV/decade limited by its operation mechanism. (Panel b adapted with permission from Ionescu et al. [4])

## III. Field Effect in Anderson and Pierels Insulators

### A. Field effect in Anderson insulators

A first step toward building a transistor is to realize field-effect in a homogeneous material. One could think of Anderson insulators as a channel layer in FET. External gate voltage could modulate the channel carrier densities and thus the Fermi level of the system. If the Fermi level could be dynamically tuned above or below the mobility edge, it is possible to induce a metal-insulator transition in the channel. There have only been a few reports on the field effect in Anderson insulators.

Figure 7(a)[89] shows field effect in a highly-disordered amorphous indium oxide film at 1.3 K with glass substrate as back gate. Amorphous and polycrystalline $In_2O_3$ could be either a weakly or strongly-disordered Anderson insulator based on its microscopic structure.[90] Interestingly, the conductance increases with both gate polarities and is almost symmetric about 0V gate voltage. Such ambipolar field effect only happens for films that have resistance larger than a certain value. For mildly disordered $In_2O_3$ films, the change in conductance is proportional to $V_G$, which is typical for a conventional FET. Furthermore, such anomalous phenomena only happen at low temperatures and the field effect conductance modulation becomes normal at high temperatures. Later several experiments[91] found that there is a relaxation process of the channel resistance after applying a gate voltage as well as a 'memory' effect that memorizes the history of the device as summarized in the following: when quenching to a low temperature and applying a constant voltage $V_G^0$ during the cooling, there will be a slow and almost 'endless' decrease in the conductance as shown in Figure 7(b). The corresponding conductance behavior does not come from the charging of the gate capacitance because the channel current relaxes much faster. Then if the field effect conductance modulation is measured, for example, by sweeping the gate voltage at certain sweep rate, the conductance versus gate voltage curves will always show a cusp similar to Figure 7(a).[91] The dip (smallest conductance) would be at the voltage $V_G^0$, where the system is cooled down and grows deeper and deeper with time. Figure 7(c) and (d) show the 'two-dip' experiment in gated Anderson insulators. The system was first equilibrated under gate voltage $V_G^0$ for ~ 24 hours at a fixed temperature. Then the gate voltage was shifted to $V_G^n$ and kept constant during later experiments. At various times after the shift to $V_G^n$, gate voltage is rapidly swept through a voltage domain and the field effect modulation of conductance will show two dips centered at $V_G^0$ and $V_G^n$, respectively as shown in Figure 7(c) and (d)[91]. Thus, the system retains the memory of both gate voltages.

These phenomena seem to be quite universal for strongly disordered systems, such as amorphous and polycrystalline $In_2O_3$, Al granular[92] separated by $Al_2O_3$ and disordered metals[93], and may be a common feature of Anderson insulators based FET. Furthermore, the ambipolar gating effect seems to be stable for a specific sample at a given temperature, and the normalized cusp shape does not strongly depend on change in structural disorder or external magnetic field[91]. The above phenomena may be related to the non-equilibrium behavior of Anderson insulators. It has been suggested that Anderson insulators with a non-equilibrium distribution of electrons always have larger conductivity than equilibrated state.[94] Upon sweeping or switching gate voltage, the system is out of equilibrium with injected or extracted electrons, which is ~1% change in carrier concentration. The comparatively small change in carrier density could cause quite significant non-equilibrium effects[89] and the dynamics of conductance modulation is actually a reflection of the process of reaching equilibrium. Although this argument is insufficient to explain all the memory effects, it may provide some insights into the operation of Anderson insulator-based FET that the device is always in non-equilibrium in practice and the realization of tuning Fermi level above and below mobility edge is difficult. The above phenomena have many features similar to glass transition and are sometimes referred to as ‘electron glass’.[95-99]

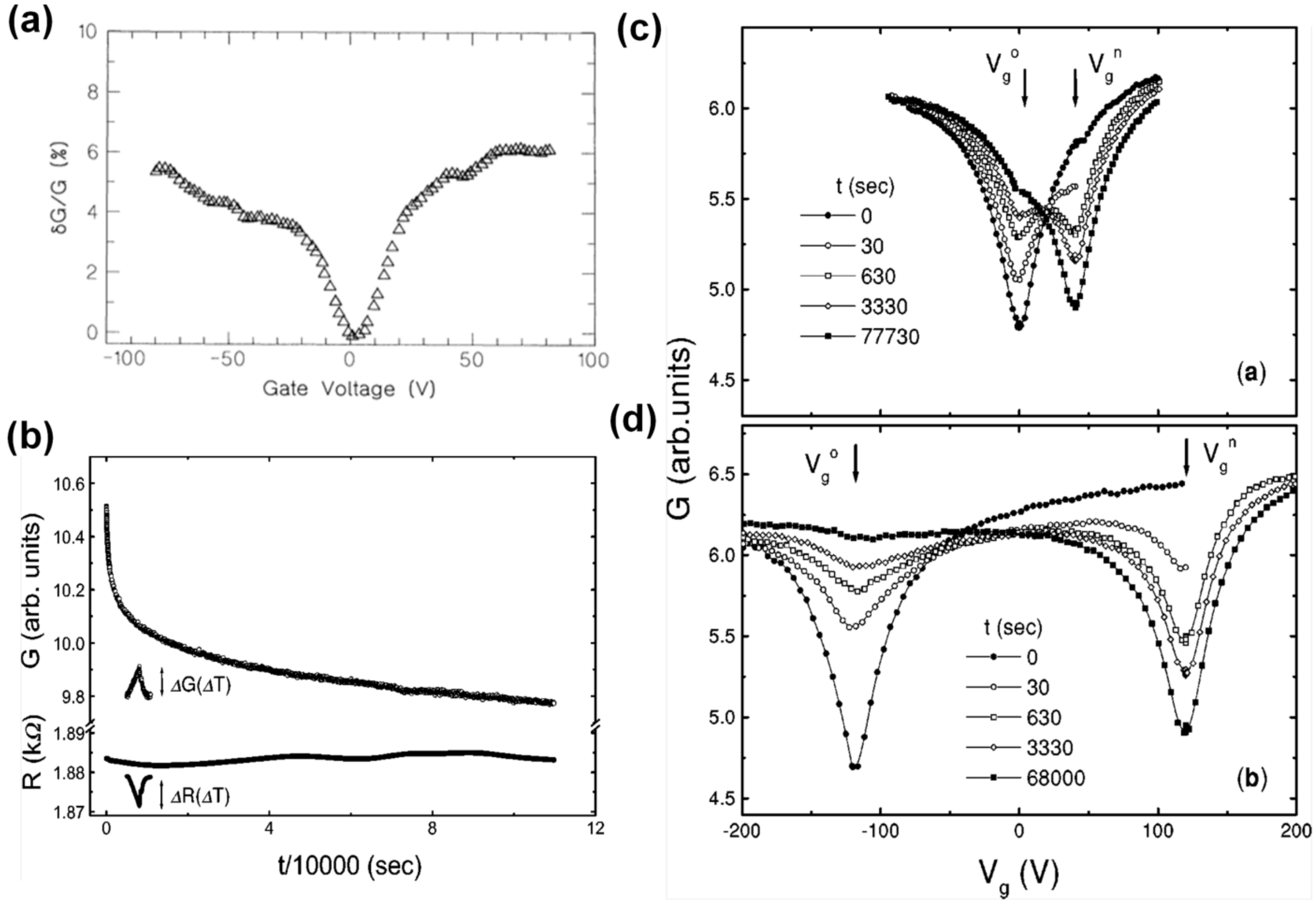

Figure 7. (a) The field-effect for strongly disordered amorphous $In_2O_{3-x}$ thin film measured at 1.3K. The channel conductance shows ambipolar increase on both voltage polarities. (b) The dynamics of channel conductance *G* when quenching the sample from ~100K to 4.1K. The lower panel showing the resistance of the Ge thermometer *R* indicates that there is no fluctuation in temperature. The conductance continues to decrease for a long time after quenching. The response of *G* and *R* to a temperature change ~1mK is also shown and could not account for the slow dynamics. (c) and (d) ‘Two-dip’ measurements in Anderson FET. The system was equilibrated under first $V_G^0$ and then $V_G^n$ for a long time. A rapid sweep in gate voltage shows that there are two dips in the conductance-gate voltage curve. Both dips grow deeper as time elapses. The system keeps a ‘memory’ of gate voltage history. (Panel a adapted with permission from Ben-Chorin et al. [89]; Panels b to d adapted with permission from Vaknin et al. [91])

## B. Gating effect in Peierls insulators

There do not seem to be any reports on electric field induced Peierls transition and could be challenging, because the transition is not directly related to charge densities and ideally a CDW could form at any band-filling in Peierls insulators. However, the electric field could alter CDW pinning conditions and consequently modulate its conduction near threshold $E_T$. Figure 8 shows the electric field induced conduction modulation in a three terminal device with a CDW conductor, $NbSe_3$, as channel material on the back gate $SiO_2$ of 55nm thickness at 30K.[100] $NbSe_3$ goes through two Peierls transitions at 145K and 59K, but still has a small part of Fermi surface ungapped below 59K, remaining metallic with carrier density of ~ $6\times10^{18}$ $cm^{-3}$. The change of current-voltage slope at ~ ±5 V corresponds to de-pinning of CDW upon threshold. The field effect modulation of single-particle conduction below threshold is about 0.1% and not visible in the figure. The modulation of differential resistance above the threshold is also rather small, but gate voltage shows a large modulation of threshold voltage. If the source-drain is biased near the threshold, a gate bias could turn on/off the CDW conduction. The mechanism of threshold voltage modulation is not clear and may be related to CDW's transverse variation or CDW gap energy. Another Peierls insulator, $TaS_3$, also shows similar response to gate bias.[100]

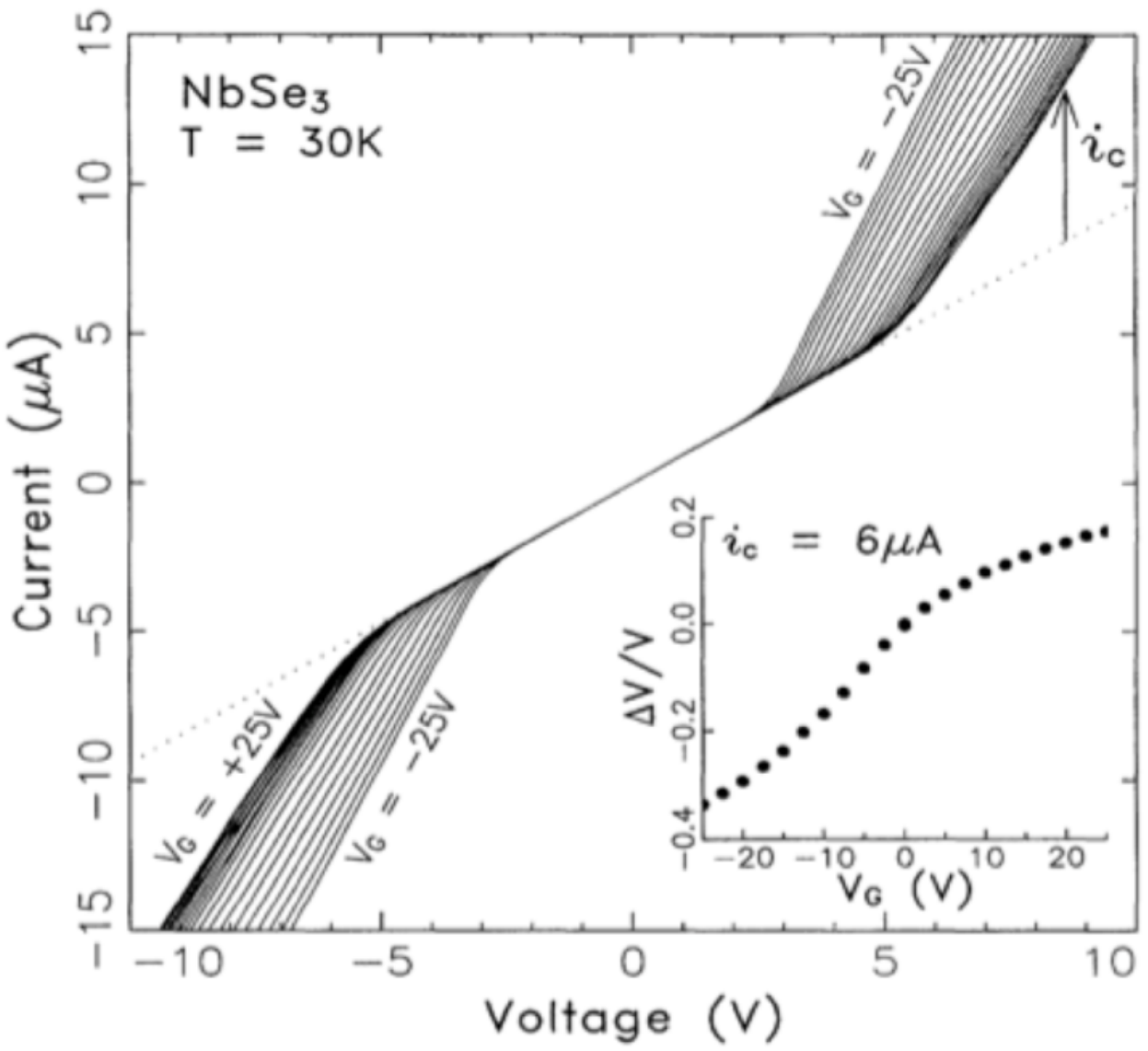

Figure 8. Field-effect in a Peierls insulator $NbSe_3$ at 30K. The field effect modulation of single-particle conduction below threshold is very small due to large carrier density in $NbSe_3$. But there is an obvious gate voltage-modulation of the threshold voltage for charge-density-wave conduction. When the source-drain is biased near the threshold, a gate bias could turn on/off CDW conduction. (Adapted with permission from Adelman et al. [100])

## IV. Mott FET – Theoretical Description

### A. Operation mechanism

Figure 9(a) shows schematic structure of proposed Mott field-effect transistor (Mott FET). Unlike conventional MOSFET devices with *n-p-n* or *p-n-p* doping in source, channel and drain area, Mott field-effect transistors could be fabricated on a homogeneous material without spatially varying dopant profiles. In the following discussion, we mainly focus on Mott field-effect devices with undoped Mott insulators as channel.

The operation of Mott FET is based on electrostatic control of carrier densities in the channel area and hence inducing electronic phase transitions of Mott insulators.[101] At finite bias, there will be carrier accumulation in the Mott insulators at the interface. Under a large enough positive or negative bias, when the electrostatic modification of charge density is sufficient to induce a Mott type insulator to metallic phase transition, there will be a large increase in the channel conductance and the device is turned ON at this threshold voltage.

In the OFF state, the density of states is zero around the Fermi level of the Mott insulator, while it becomes finite in the ON state above threshold voltage. However, it is not very well understood how the band structures or the density of states evolves with the varying doping level in Mott insulators, which makes it formidable to describe its FET behavior in the subthreshold region.

There are possibly two ways how the density of states evolves with varying doping concentration. One is to develop finite density of states at the Fermi level while the Fermi level is fixed as shown in Figure 9(b).[102] The developed states at the Fermi level are called "midgap states".[103] A second way is that the chemical potential shifts towards conduction (valence) band upon electron (hole) doping and there will be a spectral weight transfer from lower (upper) Hubbard band to upper (lower) Hubbard band as shown in Figure 9(c).[104] Spectral weight transfer could be understood in the following picture: considering an undoped half-filled Mott insulator composed of $N$ sites and one electron per site, when we dope the material with $m$ holes, the number of singly occupied states becomes $N$-$m$. There are now $N$-$m$ ways to add an electron to make a doubly occupied site, and therefore the total density of states of upper Hubbard band (related to the doubly occupied states) would be $N$-$m$. On the other hand, the total density of states of lower Hubbard band becomes $N$+$m$, because $2m$ states from the $m$ unoccupied sites now have the energy scales of lower Hubbard band and contributes to the density of states. On the contrary, in a normal semiconductor, the conduction band and valence band is rigid against doping and thus the total density of states of each band does not change. The spectral weight transfer phenomenon in correlated electron systems is distinct from uncorrelated systems[105,106] and therefore is an experimental signature of electron correlation.[107-109]

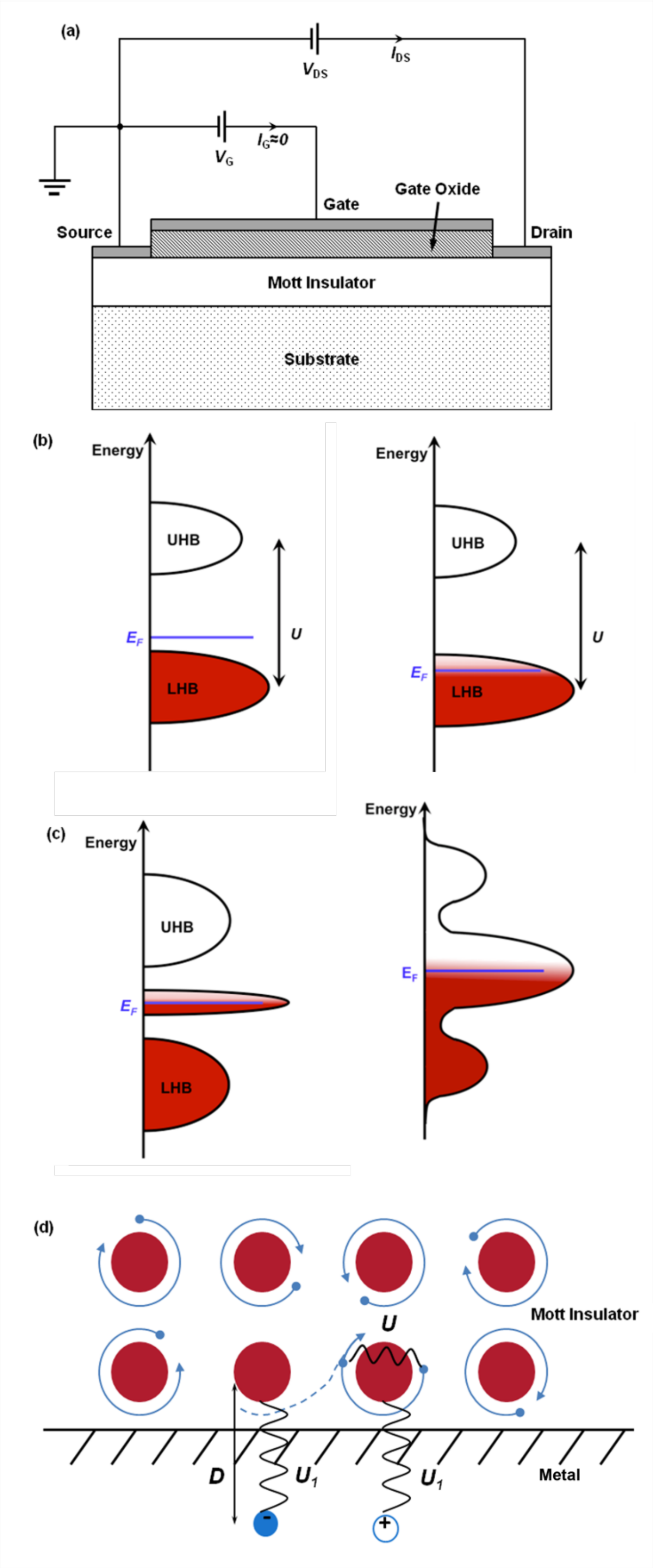

Figure 9. (a) Schematics of the proposed Mott field-effect transistor. Channel, source and drain are fabricated from a uniformly doped/undoped Mott insulator. A gate insulator could inject electrons/holes into the channel. When the injected carrier density is large enough to induce a insulator to metal transition, there is an increase in the source-drain current. (b)-(c) How the carrier density changes with surface potential in correlated systems is crucial to understand the subthreshold behavior of Mott FET. The evolution of band structure upon

doping is not well understood in these systems and two possible ways are shown. (b) Fermi level is fixed while 'mid-gap' states developed upon doping. (c) A spectral weight transfer happens accompanied by a shift in the chemical potential upon doping. The transferred weight is directly proportional to the doped carrier density. (d) When a Mott insulator is brought into contact with a metal, the hopping of electrons from one site to another will create image charges in the metal. The attraction between net charges in Mott insulator and image charges in the metal could lead to a decrease in the total hopping energy cost and effectively reduce the Hubbard *U*. (Panel b adapted with permission from Allen et al. [102]; Panel c adapted with permission from Veenendaal et al. [104]; Panel d adapted with permission from Altieri et al. [119])

## B. Screening length

To reduce the complexity, we first assume that the parameters of the Mott insulating channel, such as carrier density $n$, Coulomb repulsion $U$ and bandwidth $W$ etc., do NOT change at the dielectric/channel interface. Since the band structure of a Mott insulator is not rigid upon carrier doping, the density of states (DOS) profile is related to the location of Fermi level and therefore its surface potential. As a result, the free electron/hole density in the channel may not be described by $n \propto e^{(E_F - E_C)/k_B T}$ even for the non-degenerate situation, where $E_F$ is the Fermi level of the material and $E_C$ is the bottom of conduction band. Instead we use a general formula $n = n(E_F)$, $p = p(E_F)$ to describe the relation and this is where the new physical principle comes into play. For Mott insulators with electrons as majority carrier, the charge compressibility $\chi_c$ could be introduced as

$$\chi_c = \frac{\partial n}{\partial E_F} \qquad (8)$$

The compressibility is vanishing in the Mott insulating state at 0K, while it becomes finite in the gapless charge excitations in metallic state.

We can study how the surface potential $\Psi_s$ modifies the carrier density in the Mott insulator, which would provide information on screening length and Mott field effect transistor performance. The exact solution of carrier distribution would require the solution of Schrodinger-Poisson equation.[110] Typically, the spatial extent of the charge distribution (screening length) in the channel is larger when calculated quantum mechanically than classically, due to the electron wave-like properties.[110] However, some theoretical studies have found that the interface electron density is well captured by Poisson equation considering the lower (upper) Hubbard band as valence (conduction) band in conventional band bending picture.[111] Consequently, starting from Poisson equation and keeping the first order term, we can get the screening length in a Mott insulator would be[112-114]

$$L_D = \left( \frac{\varepsilon_r \varepsilon_0}{e^2 \chi_c} \right)^{\frac{1}{2}} . \qquad (9)$$

As previously mentioned, the charge compressibility $\chi_c$ in correlated electron systems is not well understood. The compressibility could vary strongly at the boundary of the phase transition because of the change in the DoS. For a Mott insulator at 0K, the charge compressibility is zero and the screening length is very large, which implies that the Mott insulator acts as a dielectric. At finite temperatures, the charge compressibility is non-zero in the Mott insulating state because of the thermal excitation of the free carriers and the channel has a finite screening length.

To the zeroth order, we could assume that the DOS does not change upon doping and use the formula for conventional semiconductor (equation (5)) as estimation for the screening length. As the carrier concentration is often comparatively large in correlated electron systems, the screening length would be quite small (~nm) from equation (5), which requires the channel surface to be atomically flat. From the zeroth order assumption, the material that could be metallized would have to be quite thin, making the transition characteristics almost two-dimensional.

**C. Subthreshold behavior**

As mentioned previously, metal-insulator transitions could be either mass diverging type (mainly influencing mobility) as in the Brinkman-Rice picture or carrier density vanishing type (mainly changing free carrier density) as in the Mott-Hubbard picture. As a result, there seem no general rules how carrier mobility changes as the transition happens. Experimentally, however, it is found that for many material systems, such as $VO_2$, $NdNiO_3$ and $Ni_{1-x}S$, the mobility does not change much as the carrier density across the phase transition, both for thermally induced and electrostatically induced metal-insulator transitions.[77, 170, 171, 192] Therefore, we assume in the following discussion that the transconductance behavior in the subthreshold region is dominated by the change in carrier density.

In a conventional semiconductor FET where free carrier density in the channel is the same as the net carrier induced by the gate oxide and is directly related to gate capacitance, the free carrier density in Mott FET is different from the net carrier density injection due to possible band structure changing or even band closing. The above charge density $\rho(x)$ gives the distribution of net carrier density. For example, supposing two Hubbard bands are merged in a originally half-filling Mott insulator by injecting net charge $Q_{net}$ of ~15% net carriers per unit cell, the free carrier density now becomes ~1 per unit cell, ~10 fold of the injected carriers. In other words, the free carrier density is enhanced in a Mott FET and we can introduce a parameter K, which is the ratio of free electrons to the net electrons in the channel to describe the charge enhancement.

The channel conductance under a certain gate voltage is determined by the free carrier density in the channel. Thus the subthreshold swing is

$$S = \frac{dV_G}{d\Psi_s}\frac{d\Psi_s}{d(\log_{10} I_D)} = \left(1+\frac{C_{Mott}}{C_{ox}}\right)\frac{d\Psi_s}{d(\log_{10} n)} \quad (10)$$

where n is the free carrier density, $C_{ox}$ is the capacitance of the gate dielectric and $C_{Mott}$ is the effective capacitance of the Mott insulating channel. The relative magnitude of the capacitance of the gate dielectric and channel will determine how much gate voltage will drop on the Mott insulating channel. The effective capacitance of the Mott channel should include the contribution from both its dielectric polarization and screening layer. It has been previously noted that the capacitance of correlated electron systems is related again to the charge compressibility of the material due to the screening of electric field.[211] Because the charge compressibility could vary during the phase transition, the body factor may not be a constant during the electrostatically induced phase transition. A discussion on the capacitance of correlated systems could be found elsewhere.[211] In the OFF state, $n = n_{net}$, while $n > n_{net}$ above the threshold. If we assume the net carrier density changes with surface potential in a similar way as the conventional semiconductor whereas the free carrier density is enhanced by K times with respect to net carrier density, the average subthreshold swing would be decreased by $1/\log_{10}(K)$ in Mott FET. Also it has been estimated that the cutoff frequency could be enhanced by a similar factor of K in a Mott FET over that of a MOSFET.[113]

### D. Interface properties

The channel/gate interface properties such as interface states are crucial in the operation of conventional MOSFET. In Mott FET, interfacial properties could be of even greater importance and complexity because the rich phases of Mott insulators coupled with their electron density. Understanding what would happen when the gate dielectric (often a band insulator) and the Mott insulating channel are brought into contact is a non-trivial problem. First, there will be electron redistribution due to different work function and this will build up a potential at the dielectric/channel interface. The carriers transferred in Mott insulators may induce a change in the band structure or a phase transition in the channel.[115,116]

Besides the change in carrier density and related phenomena at the interface, one would also expect the change of parameters such as bandwidth $W$ and Hubbard $U$ near the interface in Mott insulator with fixed band-filling. The reduction of Hubbard $U$ in proximity to a metal has been predicted theoretically[117] as well as observed in experiments.[118,119] As can been in Figure 9 (d)[119], the creation of a positive charge in a Mott insulator will simultaneously induce a negative image charge in the metal. The attraction between the net charge in the Mott insulator and the image charge could reduce the energy cost for electron hopping by $2U_1 = e^2 / 2\pi\varepsilon D$, where $D$ is the distance between the electron and image charge, $\varepsilon$ is the dielectric constant of the correlated material. The effective Hubbard $U'$ would become $U' = U_0 - 2U_1$, where $U_0$ is the bulk value. Note that the reduction in $U$ is purely surface phenomenon and does not involve charge redistribution between metal and correlated insulator. Similarly, one could envision a change in Hubbard $U$ when a Mott insulator is put in contact with a band insulator, where the difference in dielectric constant between two materials will also induce an image charge.

Also there may be spatial inhomogeneity that arises from phase separation in the channel. If a metallic phase is created amid an insulating matrix, for example, the channel may still appear insulating if the metallic regions are not connected.

An interesting aspect of Mott FET is that potentially it could be turned ON in both positive and negative gate polarities, which is similar to conventional bipolar FET with homogeneous doped channel (from hole to electron conducting). As discussed previously, either hole doping or electron doping could metallize a Mott insulator, but the critical density for hole and electron doping may be asymmetric.[120] Consequently one would expect different threshold voltages for positive (electron accumulation in channel) or negative (hole accumulation) gate bias. Besides different critical densities, electron or hole doping may also induce novel phases that do not appear on the other doping side, like pseudogap phase in high $T_c$ superconductors. What's more, chemically doping electrons and holes naturally need different dopants in the Mott insulators, which may give rise to different crystal structure or lattice strain upon hole or electron doping. For example, at present, $Y_{1-z}La_z(Ba_{1-x}La_x)_2Cu_3O_y$ (YLBLCO) is the only cuprate compound that could be doped on both sides without crystallographic structural change[121 122], whereas the maximum doping concentration is rather limited.[123] As a result, electrostatic doping a Mott insulator in FET structure could potentially facilitate the understandings of strongly correlated electron systems in a clean and sophisticated way without introducing disorder.

## V. Mott FET – Experiments

### A. Solid dielectric gated Mott FET

#### 1. Correlated oxide FET

Figure 10(a) shows the device structure of an early attempt[124] to build Mott FET with $Y_{1-x}Pr_xBa_2Cu_3O_{7-\delta}$ (YPBCO) as channel material. The doped n-type strontium titanate $SrTiO_3$ substrate forms the gate electrode and a layer of 400 nm thick undoped $SrTiO_3$ grown on top serves as the gate oxide. Because of the high dielectric constant ($\varepsilon_r \sim 10^2 - 10^4$ near room temperature) of $SrTiO_3$[125,126] and the ability to grow perovskite oxides epitaxial on $SrTiO_3$ single crystal substrate, $SrTiO_3$ is a popular candidate gate oxide. In addtion the dielectric constant of $SrTiO_3$ is electric-field dependent[127] and this provides another platform for device functionality.[128] After the undoped $SrTiO_3$ is deposited, the YPBCO channel is epitaxially grown by pulsed laser ablation process. Platinum electrodes are deposited as source and drain electrodes and finally device is isolated by an isolation trench. In undoped cuprates, the electron configuration of Cu is $d^9$ and it is insulating due to electron correlation. About 15% of hole doping is needed to induce a transition from insulating state to metallic state. The stoichiometry of YPBCO is designed so that it lies near the boundary of metal-insulator transition and therefore a modest change in carrier density may induce a phase transition. Figure 10(b) shows the source-drain current under different gate voltage. The source-drain current increased by about $10^2$ under a bias of -12 V while the conductance change under positive bias is almost negligible[129], which is consistent with the p-type conduction in YPBCO. The comparatively large source-drain current comes from the upper inactive cuprate part because the cuprate thickness is much larger than the screening length (~1 nm). The channel conductance increases quadratically with gate bias above threshold for the Mott FET[129], different from conventional MOSFETs where the conductance increases linearly with $V_G$. The transconductance increases linearly with gate voltage in the Mott FET and the transconductance value reaches 2 μS at a $V_D$ = 1 V and $V_G$ = 10 V. Such non-linear increase in channel conductivity upon hole doping is argued to be a result of the nonlinear increase in mobility rather than in carrier concentration. However, some ambiguity remains since voltage-dependent MOS capacitance or Hall measurements were not reported. Another concern is comparatively low field effect mobility measured, which is about 0.1 $cm^2/Vs$.[130] Field-effect experiments have also been done in cuprate $NdBa_2Cu_3O_{7-\delta}$ (NBCO) using $SrTiO_3$ single crystal bottom gate of 110 μm thickness.[131] The maximum change in resistance is 37% for gate bias -250 V and 200 V, while operating at ~50 K. In these bottom-gate FET designs, the cuprate channel is grown after the deposition of the gate oxide on the substrate, so the channel/gate oxide interface may not be atomically flat. Because FET transport properties are dominated by the free carriers within the screening length, the interface roughness will lead to degradation of transconductance.[132] Also the channel surface is subject to moisture contamination in the back gate design.[132]

To overcome these disadvantages, a new Mott FET design using buried channel was developed. In this design, the channel material is sandwiched between undoped $SrTiO_3$ substrate and an undoped $SrTiO_3$ top-gate. One of the challenges of using YPBCO and YBCO as buried channel materials is that their contact resistance with electrode is too large (~MΩ) to make the application practical.[133] On the other hand, the contact resistance in another cuprate Mott insulator, $La_2CuO_4$ (LCO)[134], is significantly smaller. A buried channel device using LCO achieves a field-effect transconductance of 45 μS at $V_D$ = 1 V and $V_G$ = 2 V.[132] The maximum $\Delta R/R$ observed is 240% with the maximum induced surface charge density equaling about $5\times10^{13}$ $cm^{-2}$. Using impedance spectroscopy technique, it was identified that the LCO grown on $SrTiO_3$ has a capacitance component coming from the grain boundaries, which may reduce the performance of the transistor. Using a better lattice matched substrate such as $LaAlO_3$ helps further improve the mobility of LCO.[130]

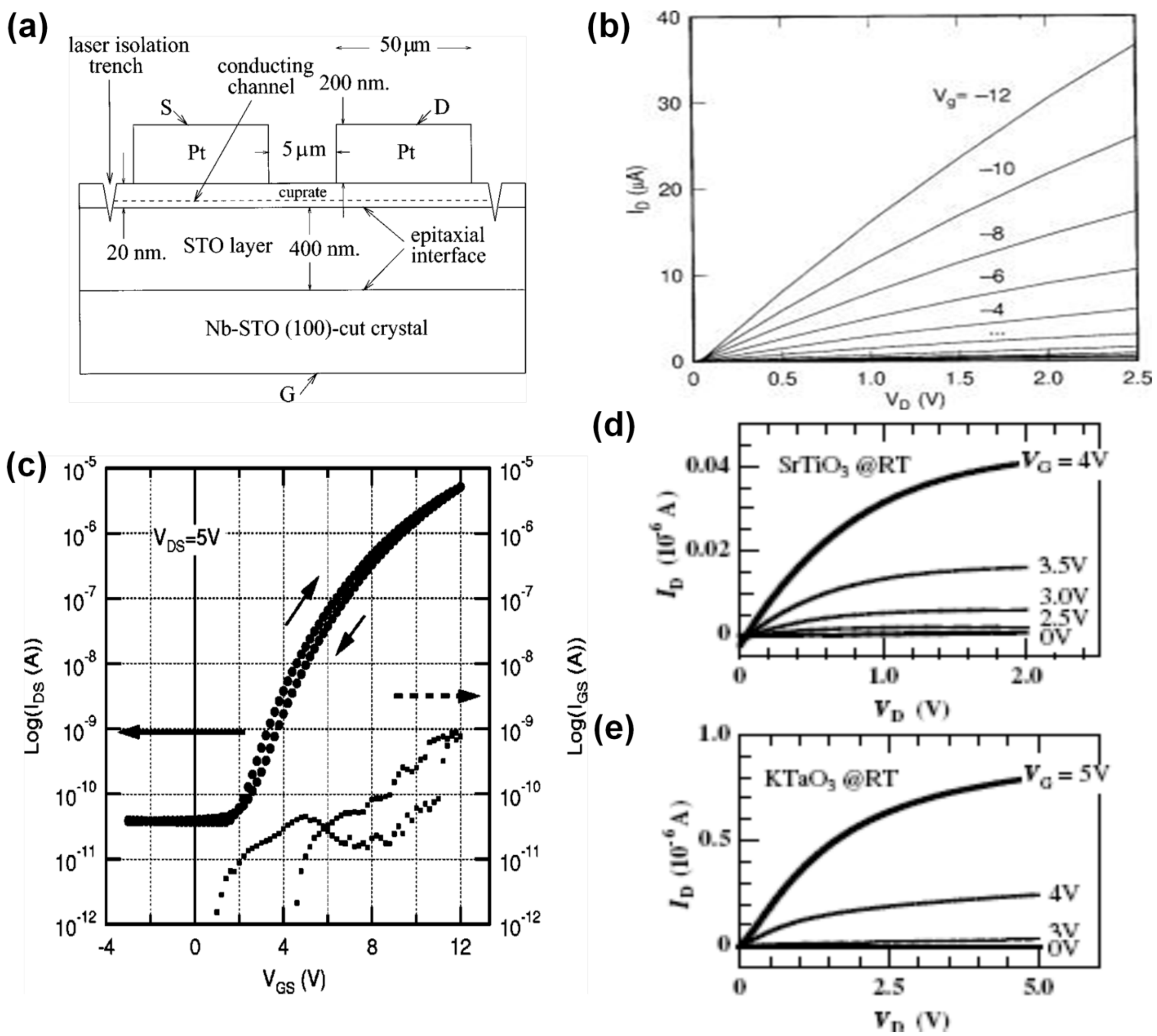

Figure 10. (a) The structure of a fabricated Mott FET with $Y_{1-x}Pr_xBa_2Cu_3O_{7-\delta}$ (YPBCO) as channel material[124]. The doped strontium titanate $SrTiO_3$ substrate formed the gate electrode and a layer of undoped $SrTiO_3$ served as the gate oxide. (b) The source-drain current of YPBCO FET under different gate voltage[124]. The source-drain current increased by about $10^2$ under a bias of -12V. (c) The channel current and the gate-source leakage current under gate voltage sweeps in a $SrTiO_3$ channel device using $CaHfO_3$ gate[139]. The field-effect mobility was 0.43 $cm^2$/Vs with on-off ratio ~$10^5$. (d) and (e) A comparison between $SrTiO_3$ channel (d) and $KTaO_3$ channel (e) FET gated by $Al_2O_3$ fabricated by same technique[147]. $KTaO_3$ based FET shows better quality due to higher carrier mobility. (Panels a and b adapted with permission from Newns et al. [124]; Panel c adapted with permission from Shibuya et al. [139]; Panels b to d adapted with permission from Ueno et al. [147])

Besides being used as gate oxide in FET designs, $SrTiO_3$ has also been used as channel material in Mott FET. Undoped $SrTiO_3$ is a band insulator with a band gap of 3.2 eV in cubic perovskite structure, whose Fermi level is close to the bottom of the conduction band. Undoped $SrTiO_3$ is a regular semiconductor with $d^0$ electron configuration. When doped with electrons into the 3d band of Ti (which is often done by substituting Sr with La), however, $SrTiO_3$ transforms into a correlated metal and the electron correlation effect becomes important.[37] With chemical doping of $10^{18}$ electrons/$cm^3$ and $10^{19}$ electrons/$cm^3$, $SrTiO_3$ becomes a correlated metal and a superconductor, respectively.[135,136] When Sr is completely substituted by La, it becomes a Mott-Hubbard insulator. As a result, $SrTiO_3$ goes from a band insulator to a correlated metal and finally to a Mott insulator as electrons are doped. Therefore, the operation of $SrTiO_3$ field effect transistor could be thought as n-type accumulation oxide based FET, but electron correlation effect also may play an important role, especially for large carrier accumulation density. Besides chemical doping, recent experiments have shown that the superconducting phase could be induced by depositing a few layers of $LaAlO_3$ (a Mott insulator) on top of $SrTiO_3$, which is related to charge transfer between the two oxides.[137] In general, the relative small number of carrier modulation needed to induce a phase transition makes $SrTiO_3$ a candidate to build FET with high ON/OFF ratio. Using 50nm thick amorphous $Al_2O_3$ sputtered at room temperature on the top of $SrTiO_3$ single crystal as gate oxide, accumulation type of $SrTiO_3$ FET was demonstrated.[138] The on/off ratio exceeds 100 with gate voltage 4 V and the threshold voltage is about 1.5 V at room temperature. However, the measured field effect mobility is only $10^{-4}$ $cm^2$/Vs. Utilizing 55nm thick amorphous $CaHfO_3$ as top gate, ON/OFF ratio of ~$10^5$ was observed with 15V gate voltage also at room temperature.[139] Figure 10(c) shows the channel current and the gate-source leakage current under gate voltage sweeps on the device using $CaHfO_3$ gate. The leakage current is orders of magnitude smaller than the source-drain current and the measured FE mobility is about 0.5 $cm^2$/Vs. The maximum carrier density accumulation is similar for amorphous $Al_2O_3$ and amorphous $CaHfO_3$ gate.[140,141] The improved field effect properties of $CaHfO_3$ gate over $Al_2O_3$ may be related to its better lattice match various perovskites and thermal stability.[141] The FET behavior is further improved by growing $CaHfO_3$ layer on $SrTiO_3$ epitaxially and a field effect mobility of around 2 $cm^2$/Vs was demonstrated at room temperature.[142] Furthermore, electric field induced metal-insulator transition is inferred from the temperature dependence of channel resistance under various gate bias[143]. The channel resistance shows thermal-activated temperature-dependence below $V_G$ = 1.3V and the activation energy reduces to zero at the crossover 1.3V gate bias. $SrTiO_3$ FET with mobility ~2 $cm^2$/ Vs, threshold gate voltage +1.1 V and subthreshold swing ~0.3 V decade$^{-1}$ was also demonstrated using 150-nm-thick amorphous $12CaO\bullet Al_2O_3$ as gate dielectric.[33]

Besides its room temperature operation, the low temperature behavior of $SrTiO_3$ FET is also of scientific and technological interest. For example, with 0.53 μm thick parylene insulator ($\varepsilon_r$ = 3.15), $SrTiO_3$ FET shows around $10^7$ ON/OFF ratio and metallic temperature-dependent resistance behavior at low temperatures.[144] The calculated carrier mobility and field-effect mobility are similar and have a value of 1000–2000 $cm^2$/Vs at 7 K. The electrostatic doped sheet electron density of $SrTiO_3$ in the above experiments was estimated to $10^{13}$ $cm^{-2}$.[139] There have also been efforts to build Mott FET based on doped-$SrTiO_3$. For example, using a ferroelectric layer $Pb(Zr_{0.2}Ti_{0.8})O_3$ to tune the carrier concentration, 30% change in resistivity and a 20% shift of superconducting temperature $T_c$ ($\Delta T_c$~ 0.05 K) are achieved in the ferroelectric FET.[145] The modulation in $T_c$ could be mapped reasonably well into the chemically doped $SrTiO_3$ phase diagram.

The technique to fabricate $SrTiO_3$ based FET has been extended to $KTaO_3$ based FET. $KTaO_3$ is an *n*-type semiconductor band insulator ($d^0$ electron configuration) with perovskite crystal structure. It has similar band structures with $SrTiO_3$ and electron correlation becomes important when it is doped with electrons. But unlike Ti that has three stable oxidation states (2+, 3+, 4+), Ta is only stable at 5+ oxidized state, which makes it challenging to chemically dope $KTaO_3$ with an electron density larger than ~$1 \times 10^{20}$ $cm^{-3}$ (1% dopants per unit cell).[146] By depositing amorphous $Al_2O_3$ of 50 nm thickness on $KTaO_3$ using the same technique in $SrTiO_3$ FET,[138] an ON/OFF ratio of $10^4$ and a field-effect mobility ($\mu_{FE}$) of >0.4 $cm^2$/Vs with 5 V gate bias are obtained at room temperature[147] as shown in Figure 10(e) in comparison with $SrTiO_3$ FETs in Figure 10 (d). The field effect mobility is much larger than that of $Al_2O_3$ gated $SrTiO_3$ FETs[138] and similar to $CaHfO_3$ gated $SrTiO_3$ FETs. Additionally $\mu_{FE}$ is almost constant with varying temperature, which may suggest formation of metallic layer in $KTaO_3$. The highest field-effect mobility at room temperature is 8 $cm^2$/Vs with an ON-OFF ratio of $10^5$ and subthreshold swing of 1.2V $decade^{-1}$ using 200-nm-thick amorphous $12CaO \bullet 7Al_2O_3$ as gate.[148]

There have also been efforts to realize Mott FETs with vanadium dioxide ($VO_2$) as channel material. Vanadium dioxide goes through a metal-insulator transition at 68ºC (single crystal) with resistance change of ~four orders of magnitude. The transition is accompanied by a structural change from a low temperature monoclinic insulating phase to a high temperature rutile metallic phase. Whether the transition should be categorized as Mott[149,150] or Peierls transition[151,152], or both mechanisms are important[153] is still on-going. It has been suggested that a modest number of electrons ~ $3\times10^{18}cm^{-3}$ is needed to induce a phase transition according to Mott criterion.[154,155] Considering the screening length of about 5nm in $VO_2$ estimated from equation (5), such carrier density accumulation is likely achievable by conventional gate oxide. For that reason, the efforts to build field-effect transistors on $VO_2$ serves both as a probe to examine the physics of transition as well as attempts to realize room temperature Mott FET. A detailed review that summarizes recent efforts on this material system could be found elsewhere.[156]

### 2. Organic Mott FET

In FETs with organic Mott insulator as channel materials, ambipolar gating effect has been observed as shown in Figure 11(a)[157]. The channel is single crystal (BEDT-TTF)($F_2$TCNQ) [BEDT-TTF = bis(ethylenedithio)tetrathiafulvalene, F2TCNQ = 2.5-difluorotetracyanoquinodimethane], which is a Mott insulator with a charge-transfer gap of about 0.7 eV. The organic molecules form quasi-one-dimensional chains that are stable to Peierls and spin-Peierls transition down to 2 K. As a result, the channel remains in Mott insulating state at the temperatures (2 K - 40 K) where the field effect is measured. The transfer curve in Figure 11(a) indicates that carrier doping is asymmetric between positive and negative gate bias. The experiments serve as an evidence that both electrons and holes could be injected into the organic

Mott insulator, but the ambipolar field effect may come from transconductance effect as in normal MOSFET without band gap closing.[157]

Recent field effect experiments[158,159] on another organic Mott insulator shows favorable results toward realization of Mott FET. The channel material, an organic insulator ϰ-(BEDT-TTF)2Cu[N(CN)2]Br (ϰ-Br), was grown on $SiO_2$ bottom gate with *p*-type Si as gate electrode. ϰ-Br is a p-type Mott insulator and goes through a metal-insulator transition upon cooling[160], which is bandwidth controlled by its thermal expansion. At temperatures lower than ~10K, the material becomes superconducting. When grown on $SiO_2$, however, ϰ-Br remains an insulator down to low temperatures (<2 K) due to the lattice mismatch and stress[161]. Figure 11(b)[158] shows the temperature dependent channel resistance at various gate voltage. The conductance in the linear region could be described by thermal excitation:

$$\sigma = \sigma_0 e^{-W/k_B T} \qquad (11)$$

where $\sigma_0$ is the conductance at high temperature limit, $W$ is the excitation energy reflected by the slope of log$R$ versus $1/T$ curve. The excitation energy varies under different gate bias. $\sigma_0$ is almost identical for different gate voltage which indicates there is no macroscopic spatial inhomogeneity[162]. The transfer curves show that electron doping will increase the conductance at some threshold voltage, whereas hole doping may also increase channel conductance in some devices. This ambipolar field effect is as expected for a typical Mott insulator. Moreover, Hall measurements provide direct evidence that band gap closing has happened upon electron doping. Figure 11(b) shows the anomalous Hall coefficients in the organic Mott FET under different bias. The Hall coefficients are positive under positive gate bias and give a hole density comparable to that of metallic state at high temperatures. Furthermore, the hole carrier concentration calculated from $n = 1/eR_H$ (~$10^{14}$ $cm^{-2}$) as shown in Figure 11(c)[159] is too large to be induced by a conventional oxide dielectric. Also note that $R_H$ at low temperatures does not change much with temperature suggesting the carrier densities are not thermally activated. By assuming that both field-induced (temperature independent) and thermally activated hole carriers coexist in the channel, the carrier density and mobility could be fitted from the experimental data. The fitting results in Figure 11(d) imply that there is a sharp increase in hole density under finite electron doping (positive gate) bias and the increase of conductance above the threshold voltage is originated from increased hole mobility instead of hole density. The above observations could not be understood in the framework of a rigid band picture. Instead it suggests that the upper and lower Hubbard band merges upon a small number of electron doping and the material becomes $1/2+\delta$ filled (*p*-type) after band closing.

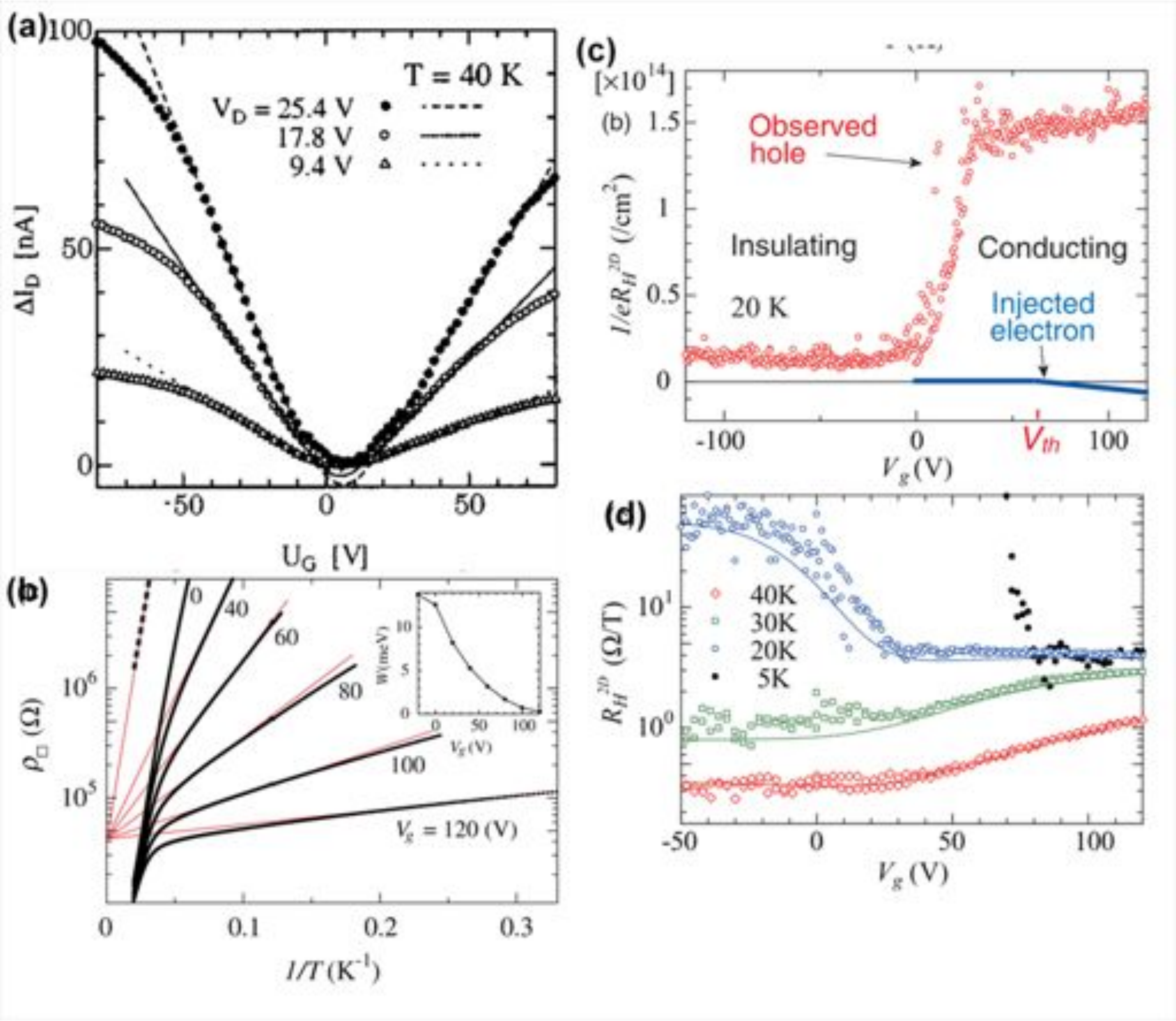

Figure 11 (a) Ambipolar transverse field-effect characteristics on an organic Mott insulator (BEDT-TTF)($F_2$TCNQ). Carrier doping is asymmetric for positive and negative gate bias[157]. (b) Arrhenius plots of sheet resistance under various gate bias in an organic Mott insulator (ϰ-Br) based FET[158]. The excitation energy *W* could be modulated by the external gate voltage. (c) Anomalous Hall effect in the organic Mott FET under different gate voltage at 20K[159]. The Hall coefficients are positive under positive gate bias and give a hole density comparable to that of metallic state at high temperatures. The hole concentration calculated from Hall measurements (~$10^{14}$ $cm^{-2}$) is too large compared with the injected electrons estimated from gate capacitance (blue line). (d) $R_H$ at low temperatures does not change much with temperature suggesting the carrier densities are not thermally activated. (Panel a adapted with permission from Hasegawa et al. [157]; Panel b adapted with permission from Kawasugi et al. [158]; Panels c and d adapted with permission from Kawasugi et al. )

There have also been reports on Mott FET based on another charge-transfer organic Mott insulator, (BEDT-TTF)(TCNQ) [TCNQ = 7,7,8,8-tetracyanoquinodimethane][30]. (BEDT-TTF)(TCNQ) goes through a metal-insulator transition at $T_{MI}$ 330K[163], which is a p-type semiconductor below $T_{MI}$ and may therefore be suitable for room temperature application. The FET device also exhibits an ambipolar field effect similar to previous experiments. Figure 11(d) shows the temperature dependence of drain-source current under various gate bias with a fixed drain-source voltage. For 80 V, 60 V, and 40 V gate voltage, there is conductance maximum at certain critical temperature ~240 K. Above the critical temperature, the resistance increases with increasing temperature, which is normally indicative of metallic behavior. The origin of this observation is not fully clarified, but the FET behavior resembles gating effect in other inorganic Mott oxides. The injected electron carrier density at 80V gate bias is estimated to be 0.014 electrons per molecule, which is normally insufficient to induce the Mott metal-insulator transition. But the small change in carrier concentration may change the transition temperature $T_{\mathrm{MI}}$, which may explain the observed phenomenon.

### B. Ionic liquid gated Mott FETs

#### 1. Gating mechanism

Conventional gate oxides could normally induce carrier density of ~$10^{13}$ cm$^{-2}$ before dielectric breakdown (1-10 MV cm$^{-1}$),[164] which corresponds to ~$10^{19}$-$10^{21}$ cm$^{-3}$ bulk density depending on the screening length. In many materials, this bulk density is insufficient to induce a phase transition such as in high $T_c$ cuprates. It has been found that a carrier density change of $10^{14}$-$10^{15}$ cm$^{-2}$ could be induced using ionic liquid (IL) as gate material in FET structures. Ionic liquid is typically a salt with organic cations and inorganic anions that is molten or in glass state in the operation temperature of FET. It is different from ionic solution in the sense that it does not contain solvents but only cations and anions. Ionic liquids usually have small electronic conductivity and comparatively large ionic conductivity. A typical structure of ionic liquid gated field effect transistor is shown in Figure 12(a). The channel surface, source and drain are covered by IL and a gate electrode isolated from channel is also in contact with IL. When a positive (negative) bias is applied to the gate electrode with respect to source/drain, cations (anions) will accumulate at the liquid/solid surface and correspondingly electrons (holes) will be electrostatically doped into the channel, provided there is no electron transfer between cations (anions) and channel, i.e. electrochemical reactions. The accumulated ions in IL forms layered structure near the interface, which is known as electric double layer (EDL). Therefore, the IL-gated FET is often referred to as electric double layer transistor (EDLT). The EDL capacitance could not be well described by Gouy-Chapman model because it is in the limit of high ion density, but one can roughly estimate its areal capacitance from a simple formula only considering the Helmholtz layer: $\frac{1}{C}=\frac{d}{\varepsilon_{IL}\varepsilon_0}+\frac{\lambda}{\varepsilon_{Ch}\varepsilon_0}$, where $d$ is the spacing between cations and anions in the Helmholtz layer, which is typically a few nanometers, $\varepsilon_{IL}$ is the dielectric constant of the ionic liquid, $\lambda$ and $\varepsilon_{Ch}$ are the screening length and dielectric constant of the channel material, respectively. The estimated capacitance is about 10 μF/cm$^2$ and such large capacitance would enable carrier accumulations of $10^{14}$ cm$^{-2}$ at ~1 V. Due to comparatively small Debye screening length in many Mott materials, the corresponding bulk density would be comparable to and may even exceed the values achievable by chemical doping.[165]

Just as in conventional solid gated FET, it is impossible to apply an infinitely large gate bias in IL gated FET. Electrochemical reactions would happen at a certain voltage and the electrochemical window (usually a few Volts, depending on the oxidization/reduction potential between ionic liquid and channel material) will limit the maximum gate voltage and hence maximum carrier accumulation in the device.

**2. Ionic liquid gated field effect experiments**

Figure 12(b)[166] shows frequency dependent $InO_x$ channel impedance in an EDLT device under different gate voltages. The semiconducting amorphous $InO_x$ films are grown on glass substrate by reactive ion beam sputtering with electron density ~$5\times10^{19}cm^{-3}$ and go through a insulator-superconducting phase transition at ~1.5K.[167] The ionic liquid used is 99.5% pure 1-ethyl-3-methylimidazolium bis(trifluoromethylsulfonyl)imide (EMI-Beti). In the high frequency limit, the IL is leaky and the total resistance is the parallel combination of ionic liquid bulk resistance and channel resistance. On the other hand, IL's capacitance is large and the total resistance mainly comes from the $InO_x$ channel in the low frequency limit. The inset of Figure 12(b) shows how the low frequency (0.02 Hz) resistance changes with gate voltage and nearly 4 orders of magnitude change in resistance is achieved. The curve is asymmetric because negative gate bias (electron depletion) is closer to metal-insulator transition than positive bias (electron accumulation). The calculated areal capacitance is 5.9 $\mu F/cm^2$ and about 1 order of magnitude larger than previous work using $Al_2O_3$ oxide gate.[167] The carrier density change estimated from capacitance agrees qualitatively with Hall measurements and is sufficient to induce a transition into insulator of $InO_x$ under $V_G = -1V$.

Similar gate voltage dependent sheet conductance and carrier density modulation has been achieved in ZnO thin films gated by $KClO_4$/polyeltheleneoxide (PEO) electrolyte.[168] Note that $KClO_4$/PEO is not an ionic liquid but instead an ionic solution with smaller areal capacitance than ILs, but its functionality in EDLT is essentially identical to ionic liquids. Figure 12(c) shows temperature dependent channel resistance from four terminal measurements under various gate voltages. Under small gate bias, the channel resistance increases with lowering temperature as an insulator while it shows a metallic temperature dependence at large positive gate bias, indicating accumulated electrons (~$3\times10^{13}$ $cm^{-2}$) drive ZnO from semiconducting state into metallic state. The fluctuation of resistance at 220 K-280 K is related to the glass transition of the electrolyte. At these temperatures, the liquid gate solidifies into glass but the induced charges remains unaltered at the gate/channel interface as long as a constant gate bias is applied. This common feature of ionic liquids enables EDLT to operate in a wide range of temperatures even below the glass transition temperature. Similar temperature dependent resistance is observed in later experiments[169] using ionic liquid, N,N-diethyl-N-(2-methoxyethyl)-N-methylammonium bis-trifluoromethylsulfonyl-imide (DEME-TFSI), achieving much higher density accumulation $4.5\times10^{14}$ $cm^{-2}$ and $5.5\times10^{14}$ $cm^{-2}$ at room temperature and down to 1.8K, respectively.

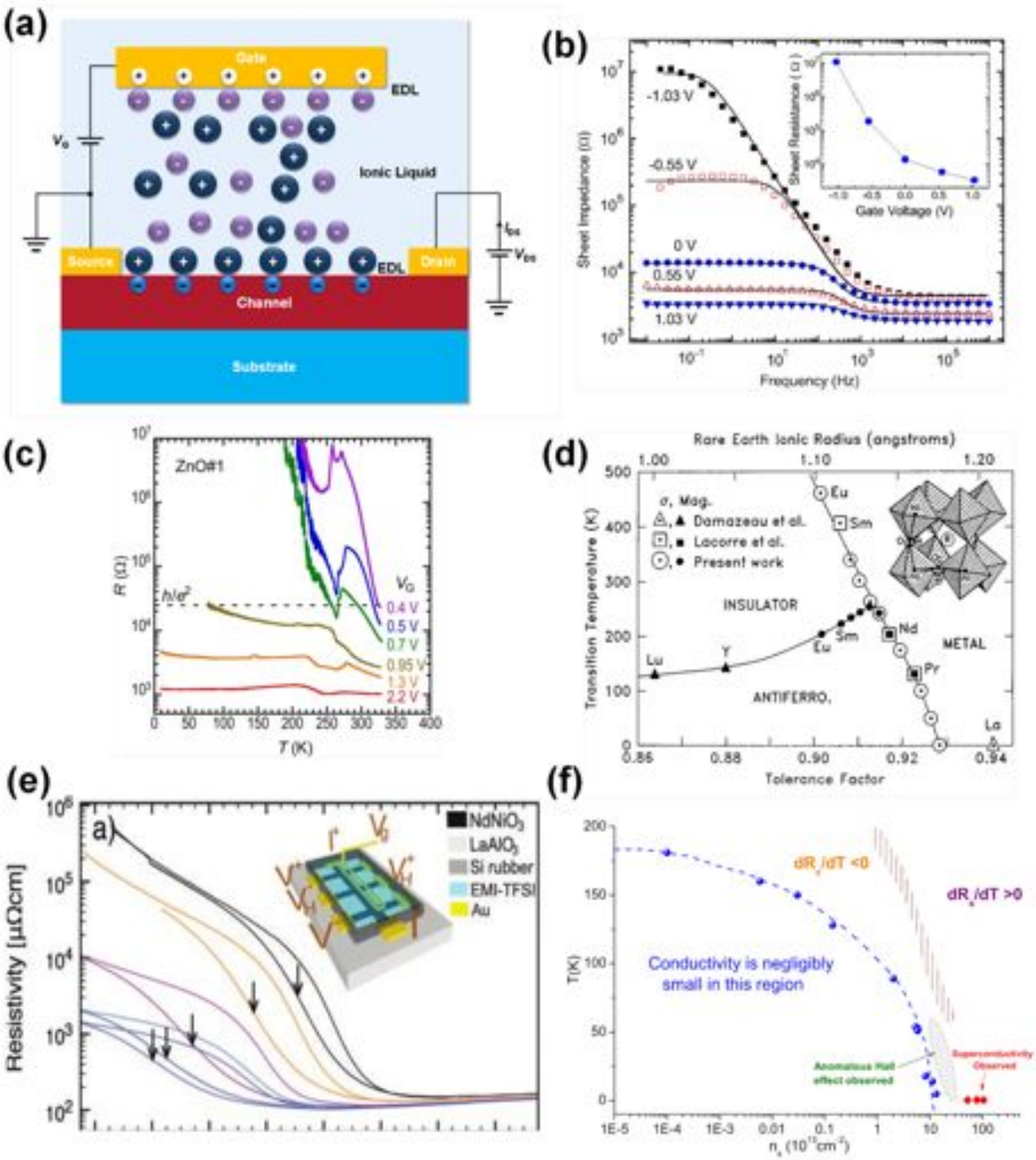

Figure 12 (a) Schematics of electric double layer transistor (EDLT). The gate oxide in conventional FET is replaced by an ionic liquid that is an electronic insulator. When a positive bias is applied to the gate electrode, cations in the ionic liquid will accumulate on the channel surface and correspondingly electrons could be electrostatically induced. The formed electric double layer typically has ultra-large capacitance and enables carrier accumulation in the channel much larger than that is achievable than conventional oxide gate. (b) The frequency dependent $InO_x$ channel impedance under different gate voltages in an EDLT[166]. The inset shows how the low frequency (0.02 Hz) resistance changes with gate voltage and nearly 4 orders of magnitude change in resistance is achieved. (c) Temperature dependent four-probe sheet resistance of ZnO at various gate voltages in EDLT[168]. The metallic temperature dependence observed at large positive gate bias indicates accumulated electrons ($\sim 3\times10^{13}$ $cm^{-2}$) drive ZnO from semiconducting state into metallic state. (d) Electronic phase diagram of $RNiO_3$.[172] (e) The resistivity versus temperature curves of $NdNiO_3$ film of different thickness gated by ionic liquid[170]. Electrostatic doping does not change the metallic conductance, but could modulate the insulating resistance by 1 order of magnitude and induce a metallic state with -2.5 V gate voltage. (f) The proposed temperature-carrier density phase diagram of $SrTiO_3$ probed by electrostatic doping with ionic liquid gating.[178] The vertically hatched region corresponds to the approximate boundary where derivative of sheet resistance R with respect to temperature changes sign – a feature of metal-insulator transition. The diamonds indicate superconductivity while the crosshatched region represents anomalous Hall effect. (Panel b adapted with permission from Misra et al. [166]; Panel c adapted with permission from Shimotani et al. [168]; Panel d adapted with permission from Torrance et al. [172]; Panel e adapted with permission from Scherwitzl et al.[171]; Panel f adapted with permission from Lee et al. [178].)

Tuning of metal-insulator transition in $NdNiO_3$, a member of orthorhombic distorted perovskite structure rare earth nickelates $RNiO_3$ ($R$ = rare earth), has been demonstrated with EDLT structures.[170,171] $RNiO_3$ series, except $LaNiO_3$ that remains metallic down to low temperature regions, undergo metal-to-insulator transitions upon cooling, accompanied by a slight but abrupt change in lattice parameters and become (charge-transfer) insulators with O *2p* as valence band and Ni *3d* upper Hubbard band as empty conduction band.[77] The transition temperatures $T_{MI}$ are related to the radius of $R$ atoms as shown in Figure 12(d)[172]. The atomic radius of the rare earth ion controls the tilting of $NiO_6$ octahedral, alters the angle between Ni-O-Ni bonding and hence *p-d* transfer interaction. This distortion results in change in the bandwidth and makes the transition in $RNiO_3$ a prototype of bandwidth control MIT. Nevertheless, substituting $R$ with $Ca^{2+}$ or $Ce^{4+}$ in $RNiO_3$ (the nominal valence state of $R$ being 3+) could also modulate $T_{MI}$, in which case band-filling controls the metal-insulator transition.[173,174] $NdNiO_3$ exhibits a metal-insulator transition from paramagnetic metal to antiferromagnetic metal at ~200 K.[172] In $Nd_{1-x}Ca_xNiO_3$ and $Nd_{1-y}Th_yNiO_3$, the metal-insulator transition temperature changes with chemical hole and electron doping by $dT_{MI}/dx$ = -3180 K and $dT_{MI}/dy$ = -1029 K, respectively.[174] This gives roughly 32 K decrease in $T_{MI}$ with hole doping of $2\times10^{20}$ holes/cm$^3$. However, doping $NdNiO_3$ with different ions also modifies the angle between Ni-O-Ni bonds due to different ion radius and alters the bandwidth $W$. Therefore similar amount of chemical electron/hole doping (for example, doping equal percent of divalent atoms $Sr^{2+}$ and $Ca^{2+}$) could lead to different modulation in the transition temperature.[174] The electrostatic doping could be a clean way to probe the electron correlations without introducing lattice distortion. Furthermore, charge order effect observed in $NdNiO_3$ may be important for understanding the nature of insulating state as well as interpreting the field effect results,[175,176] because the gate-induced charges may distribute inhomogenously in charge ordered insulators. Figure 12(e)[170] shows the resistivity versus temperature curves of $NdNiO_3$ film of different thickness gated by IL, DEME-$BF_4$ (N,N-diethyl-N-(2-methoxyethyl)-N-methylammonium tetrafluoroborate), under 0V and -2.5V gate bias. The metallic conductance is unaltered by electrostatic doping, which is similar to the results of chemical doping,[174] while the insulating resistance could be modulated by 1 order of magnitude and down to metallic state with -2.5 V gate voltage. The transition temperature is strongly reduced by hole doping, especially in ultra thin films. The change in $T_{MI}$ agrees reasonably well by the calculation from $dT_{MI}/dx$ = -3180 K/hole by estimating hole accumulation density from capacitance measurements for films of different thickness. These results indicate that the electrostatic doping behaves in a similar way as chemical doping in $NdNiO_3$. Another IL-gated experiment on $NdNiO_3$ using EMI-TFSI (1-ethyl-3-methylimidazolium-bis(trifluoromethylsulfonyl)imide) led to a resistance modulation of up to 3 orders of magnitude in the insulating state with -3V gate bias[171]. The $NdNiO_3$ films used[171] have larger thermal metal-insulator transition magnitude than the previous ones[170] and it seems crucial to grow high quality oxide films with large thermal metal-insulator transitions to achieve corresponding high ON/OFF ratio for FET devices.

Gating effect in perovskite insulators such as undoped $SrTiO_3$[177] and $KTaO_3$[34] have also been investigated in EDLT geometry. In undoped-$SrTiO_3$ based EDLT gated with $KClO_4$/POE electrolyte, areal electron density of $5\times10^{13}$ $cm^{-2}$ (corresponding bulk density ~ $7\times10^{19}$ $cm^{-3}$, about 1 order of magnitude larger than that induced by a solid gate dielectric) is accumulated at +3V gate bias and superconductivity occurs below 0.4 K.[177] Using another IL, DEME-TFSI, to gate undoped $SrTiO_3$, the temperature-carrier density phase diagram is determined by conducting Hall measurements under a series of gate voltages as plotted in Figure 12(f).[178] It is clear from the phase diagram that metal-insulator transition could be induced with $10^{13}$-$10^{14}$ $cm^{-2}$ areal electron doping. At lower temperatures and larger carrier density, superconductivity appears at ~$3.9\times10^{14}$ $cm^{-2}$ with $T_c$ = 0.4 K and $T_c$ decreases with increasing electron density. The difference in critical carrier density[177] may arise from the estimation of capacitance and the screening length.

As discussed, $KTaO_3$ FET gated by conventional oxide only shows insulator to metal transition. Due to the difficulty to chemically dope $KTaO_3$, no superconductivity was seen down to 10 mK and whether $KTaO_3$ could be doped into a superconductor remains unsettled.[179] By utilizing ionic liquid DEME-$BF_4$ (N,N-diethyl-N-(2-methoxyethyl)-N-methylammonium tetrafluoro-boron) as gate electrolyte in EDLT, the accumulated electron density around $10^{21}$ $cm^{-3}$ (~ 1 order of magnitude larger than maximum chemical doping concentration) and superconductivity was observed in $KTaO_3$. These results demonstrate the possibility of using FET structure to induce phase transitions in materials where chemical doping concentration is inadequate.

To summarize, ionic liquid gated EDLTs have broken the bottleneck of the maximum accumulated carrier concentration limited by gate breakdown in conventional FET. They have been proven to be promising to be utilized to investigate electric field induced phenomena in a broad range of materials and probe correlated electron physics. However, there seem still be technological challenges in ionic liquid gated EDLTs before they could be possibly applied outside basic scientific research. The foremost disadvantage of IL-gated EDLTs is the liquid form of gate, which will give rise to concerns on stability, scalability and integration. Another challenge is that one needs to go above the glass transition temperature of ionic liquid, $T_g$, to make it possible to modulate carrier density modulation by gate voltage, which would cause difficulties in the dynamic control and fine tuning of channel conductance at low temperatures. This is also the reason why many works using IL-gated EDLT only show a set of discrete gate voltage instead of a transfer curve. Some important FET device parameters such as subthreshold swing are not straightforward to measure in EDLT geometry. Further, IL-gated EDLTs usually have a significantly longer relaxation time compared with conventional oxide dielectric due to slow motion of ions, which may impede ultrafast switching applications. The microscopic mechanisms of slow relaxations require a more detailed understanding, as this will limit how the devices can be turned 'OFF'. The slow response also may lead to spurious observations on resistance modulations under gate bias that may have non-physical origin such as reactions/ionic migration, other. Finally, the EDLTs may not be suitable for high-temperature operation because of IL's degraded chemical stability at elevated temperatures.[180] The electrochemical reactions between IL and channel material or moisture interaction may mimic the 'pure' field effect from electrostatic doping and lead to ambiguity and complexity of data interpretation and requires a deep understanding that is presently lacking.

### C. Mott Transistors based on other gate designs

Besides using conventional oxide and ionic liquids as gate material, Mott transistors could also be designed with ferroelectric gate or be modulation-doped. We briefly provide some perspectives on how these techniques are compared with Mott FETs. In ferroelectric field effect transistors, the carrier density modulation in the channel is due to the change of the spontaneous polarization of the ferroelectric gate.[181-183] Many of the ferroelectric materials are transition metal oxides that could be grown epitaxial on oxide substrates,[184] making it easier to integrate with Mott insulators. Also the expected nonvolatile carrier density modulation may decrease the stand-by power and be utilized in memory applications. However, the channel charge densities could be only switched in discrete levels and could not be fine-tuned. It could not be utilized as a tool to probe the detailed physics of electrostatic doping as in conventional FET, but its subthershold may be superior to conventional FET for the same reason. The most serious issue about ferroelectric gate may be that the fastest switching speed of spontaneous polarization is 70 to 90 ps,[185] which is still slow compared to a state-of-the-art CMOS switch.

The heterojunction modulation-doped Mott transistor is essentially an analog of junction gated field-effect transistor (JFET) or heterostructure field-effect transistor (HFET). When a Mott insulator is brought into contact with a highly doped semiconductor, there will be electron transfer from the semiconductor to the Mott insulator if the doped semiconductor has a higher Fermi level. The Mott insulator is doped with electrons at the interface, which might induce a phase transition. Depletion layer will form in the semiconductor side and by tuning the external bias across the heterostructure, one is able to modulate the injected electron densities in the Mott insulator. By growing $NdNiO_3$ on La-doped $SrTiO_3$ with different La concentration heterostructure, the MIT transition temperature can be tuned in $NdNiO_3$, because the charge transfer varies for differently doped La concentration.[186] The dynamic modulation of channel resistance through external gate voltage, however, was not presented and only remains on a theoretical level at present.[113] Similar to the difference between JFET and MOSFET, modulation-doped Mott transistors is likely to have larger gate current than oxide gated Mott transistor and may not operate under large forward gate bias because it will lead to a decrease in the built-in potential between the semiconductor and Mott insulator and large increase in leakage current. The coupling between gate voltage and carrier density is expected to be less efficient due to smaller effective capacitance. Also it is challenging to achieve enhancement mode FETs in these devices.

## VI. Summary And Outlook

Metal-insulator transitions can be induced by disorder, electron-phonon interaction and electron correlation effects. For logic devices, strongly correlated electron systems appear promising since the phase transition onset is directly linked to the carrier density.

Table II summarizes the device parameters of a few representative correlated material FETs in comparison benchmarked with Si 32nm CMOS technology. Most of the demonstrated devices are still in early stages of research with little optimization of channel-insulator interfaces and contact resistance, and therefore have much larger device dimensions and subthreshold swing than state-of-the-art Si MOSFET. A concern with Mott FETs is that the carrier mobility in the correlated materials is usually quite low, which is related to the tight-binding nature and hopping conduction of *d* electrons.

Fabricating high quality Mott insulator/gate insulator interfaces and inducing sufficient net carrier density in the correlated material are among the most outstanding challenges towards realizing high performance FETs. The small screening length in Mott insulators requires the precise control of heterostructure growth and interfacial defects at atomic-scale. Elementary switching devices are required to improve our understanding of field effect with such materials, observe intrinsic channel properties in a domain not dominated by interfacial defects or traps and eventually dynamic characteristics. In an un-doped Mott insulator, the carrier density change induced by a conventional solid gate may not necessarily be not sufficient to trigger a phase transition. Ionic liquid gating enables large carrier accumulation density in the channel, but is not straightforward due to its large relaxation time or surface reactions. As a result, innovations in gate dielectric synthesis are absolutely necessary. It has also become crucial to find specific correlated electron systems (either doped or undoped) that lies on the verge of insulator to metal transition but also stays insulating without a gate voltage, which demands the growth of high-quality correlated materials.

To theoretically evaluate and model the device performance of correlated electron FET, some basic understanding of the Mott transition, such as how the band structure evolves with doping concentration in correlated electron systems and how do they change properties in the proximity of the interface, needs to be developed. A closely related problem is to visualize band bending phenomena for correlated electron systems. The success of classical semiconductor devices and circuits is largely due to our ability to rigorously quantify the material and device response, and develop a whole host of circuit level modeling tools that enables understanding scaling protocols. Whether such a successful counterpart for correlated materials exist remains to be uncovered, but once constructed, will be extremely useful to guide experimental work. While several works to date have reported on dc characteristics of proof-of-concept devices, ac response, dynamics, junction size effects, power dissipation per switching operation etc are largely unknown for correlated electron systems except for few analyses.[187] A systematic effort is therefore required to experimentally measure the relevant data and correlate with physical models.

In conclusion, we have surveyed the theoretical understanding and experimental efforts in the area of three-terminal field-effect devices with correlated insulators. Taking advantage of the rich physical phenomena related to tunable carrier densities in correlated electron systems, the projected correlated field-effect transistors could possibly be of interest as logic devices to compliment Si MOSFET and also enable new functionalities that could be used in new computing architecture. Additional possibilities may exist in the area of high speed switches of great interest in communication/navigation systems. Naturally, if one could integrate such new materials with CMOS technologies seamlessly, their chances of technological impact could be greater in the near-term.

**Table II.** The device parameters of demonstrated FETs based on correlated materials in comparison benchmarked with Si CMOS technology.

| Channel material | Si[188-190] | $SrTiO_3$[143] | $KTaO_3$[148] | $La_2CuO_4$[132] | $VO_2$[191,192] | (BEDT-TTF)(TCNQ)[30] |
|---|---|---|---|---|---|---|
| Channel length (nm) | 112 (gate pitch length) | 100 000 | 200 000 | 1 000 | 160 000 | ~10 000 |
| Gate material | High-k | $CaHfO_3$ | $12CaO{\bullet}7Al_2O_3$ | $SrTiO_3$ | DEME-TFSI | $SiO_2$ |
| Gate thickness (nm) | ~1 (effective) | 50 | 200 | 100 | ~1 (effective) | 300 |
| Subthreshold swing (mV/decade) | ~100 | ~250 | ~1 200 | ~10 000 | | |
| Critical carrier density ($cm^{-3}$) | | $\sim10^{18}$ | $\sim10^{19}$ | | $\sim10^{18}$ | $\sim10^{17}$-$10^{18}$ |
| Carrier mobility ($cm^2/Vs$) | ~800 | ~2 | ~8 | | ~0.1 | $\sim10^{-2}$ |